\documentclass[a4paper,10pt,twocolumn,twoside]{article}
\usepackage{amsmath, amsthm, amssymb}
\usepackage[english]{babel}
\usepackage{chicago}
\usepackage[pdftex]{graphicx}
\usepackage{hyperref}
\usepackage{color}
\usepackage[utf8]{inputenc}
\usepackage{geometry}
\hypersetup{
pdfauthor=Havemann \& Larsen,
pdftitle=Bibliometric Indicators of Young Authors in Astrophysics,
colorlinks=true,
linkcolor=black,
anchorcolor=black,
citecolor=blue,
urlcolor=black,
menucolor=black
}

\title{Bibliometric Indicators of Young Authors in Astrophysics:\\
Can Later Stars be Predicted?}
\author{Frank Havemann\footnote{Institut f\"{u}r Bibliotheks- und Informationswissenschaft, Humboldt-Universit\"at zu Berlin, D 10099 Berlin, Dorotheenstr. 26, Germany} 
\and Birger Larsen\footnote{Department of Communication,
Aalborg University, Copenhagen, Denmark}}
\date{}

\begin{document}

\maketitle
\begin{abstract}
We test 16 bibliometric indicators with respect to their validity at the level of the individual researcher by estimating their power to predict later successful researchers. We compare the indicators of a sample of astrophysics researchers who later co-authored highly cited papers before their first landmark paper with the distributions of these indicators over a random control group of young authors in astronomy and astrophysics. 
We find that field and citation-window normalisation substantially improves the predicting power of citation indicators. 
The two indicators of total influence based on citation numbers normalised with expected citation numbers are the only indicators which show differences between later stars and random authors significant on a 1\,\% level.
Indicators of paper output are not very useful to predict later stars. 
The famous $h$-index makes no difference at all between later stars and the random control group.   
\end{abstract}
\section{Introduction}

Any indicator should actually indicate what it is made for. If an indicator is used for evaluation it should not provide an incentive for an unwanted behaviour. In scholarly publishing we know salami and multiple publications, unjustified assignment of co-authorship, and different practices of tactical citation behaviour. Bibliometricians should strive to develop valid research indicators which have no unwanted adverse effects~\cite{kreiman_nine_2011}.

Most bibliometric indicators are not developed for the evaluation of individual researchers~\cite[p.\ 1565]{costas_bibliometric_2010}, however individuals are increasingly being evaluated using such indicators. We test selected indicators with respect to their validity at the level of the individual researcher by estimating their power to predict later successful researchers. For this reason, we compare bibliometric indicators of a sample of astrophysics researchers who later co-authored highly cited papers (later stars, for short) before their first landmark paper with the distributions of these indicators over a random control group of young authors in astronomy and astrophysics. 

Results obtained with some standard basic indicators have been presented on a poster at ISSI 2013.\footnote{ \textit{14th International Society of Scientometrics and Informetrics Conference} in Vienna, Austria, 15th to 20th July 2013 \cite{havemann2013cls}} Here we extend the study to more sophisticated measures with the aim to find the best indicators for predicting later stars.
We imagine that later stars apply for a job in an astrophysical research institute five years after their first paper in a journal indexed in Web of Science~(WoS). Do they perform better bibliometrically than the average of applicants with the same period of publishing?  

\section{Data and method}

\subsection{Sampling of authors}

We inspected 64 astronomy and astrophysics journals to find researchers who started publishing after 1990 and had published for a period of at least five years in WoS journals. We excluded those who had more than 50 co-authors on average because evaluating those big-science authors cannot be supported by bibliometrics. We draw a random sample of 331 authors mainly publishing in this field and affiliated longer in Europe then elsewhere. The latter criterion contradicts with the international character of astrophysics research but makes the sample more homogenous with respect to the educational and cultural background of the researchers. 

To find authors with highly cited papers, for each journal considered we ranked papers with more than four citations per year and less than ten authors according to their citations per year. We excluded papers with ten or more authors because we want to have later stars whose contributions to the successful papers are not too small. From the top 20 percent of these paper rank-lists we extracted all European authors of highly cited papers. We obtained 362 candidates who published their first highly cited paper at least five years after their first paper in one the 64 journals. 

We ranked these later-star candidates according to their number of highly cited papers. We went through this list and checked whether the authors had really five years or more to wait for the breakthrough paper if all their papers in WoS-journals are taken into account. We chose the first 40 authors to keep the effort manageable. 
For all WoS-papers of the 40 later stars and of the 331 random authors (downloaded at Humboldt-University, Berlin) all citing papers were determined by CWTS, Leiden. 
All bibliometric indicators presented below are based on papers and their citations within the first five years of the author. To compare only authors with similar collaboration behaviour we restricted both samples to authors with less than four and more than one co-author on average ending up with 30 later stars and 179 random authors. 

We further restricted both samples to authors starting before 1999 because there is only one  star starting later (in 2002) but many random authors (more than 100). By this restriction to 29 stars and 74 authors in the control group we take into account that the citation behaviour of astrophysicists has changed  remarkably during the last 25 years. 
The  numbers of references have increased. The median of reference numbers of the 448 papers published in the 1986 volume of the \textit{Monthly Notices of the Royal Astronomical Society} was 24. Till the year 2010 the median of reference numbers has doubled (calculated with 2,006 papers, data source: WoS).\footnote{ cf.\ \citeN[p.\ 5]{henneken2011ads}
} Longer lists of references induce higher citation numbers of papers. Thus, both samples still have a time variance of expected citation numbers. This time variance increases the overlap between the citation-indicator distributions of the samples when citation numbers are not normalised. In other aspects the union of our samples is surely more homogenous than many real groups of applicants (career duration, collaboration behaviour, geographical background).

An alternative data source for astrophysics publications and their citations is the \textit{Astrophysics Data System} (ADS)\footnote{ \url{http://adsabs.harvard.edu}} delivered jointly by the US National Aeronautics and Space Administration (NASA) and the Smithsonian Astrophysical Observatory \cite{henneken2011ads}. ADS includes also non-refereed publications. Any user can obtain a whole slew of bibliometric indicators for any set of selected publications. 
\newpage
\subsection{Statistics} 
\begin{figure}[!t] 
\begin{center} 
\includegraphics[width=2.8in]{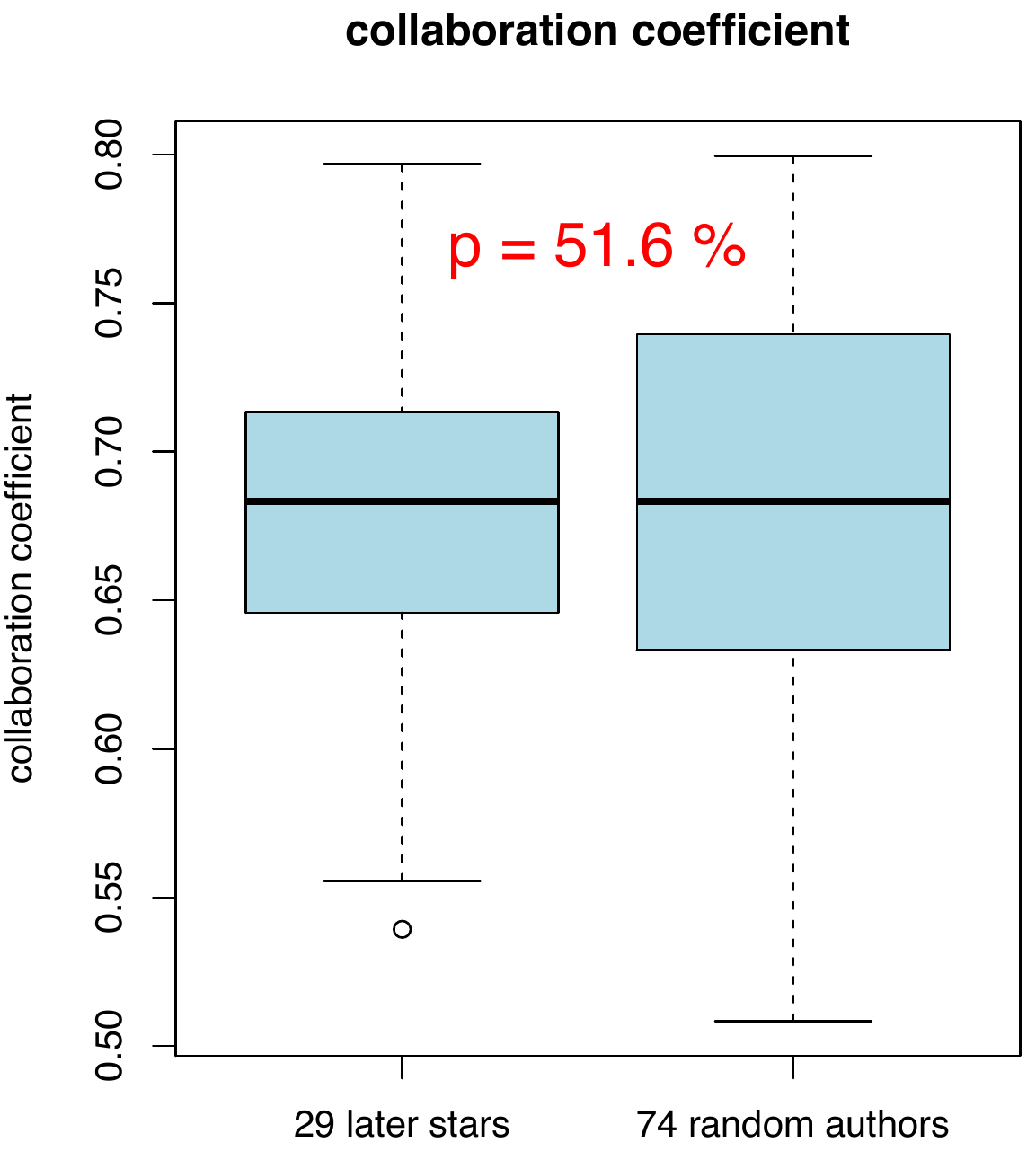} 
\end{center} 
\caption{The authors in the two samples have similar distributions of collaboration behaviour.} 
\label{Fig-coll.coeff} 
\end{figure}

For each bibliometric indicator considered, we test whether both samples behave like random samples drawn from the same population by applying a one-sided Wilcoxon rank sum test with continuity correction. We test the null hypothesis that for both samples we have the same probability of drawing an author with a larger value in the other sample. The alternative hypothesis is that indicator values of later stars exceed the values of random authors.\footnote{ cf.~the  Wikipedia article   \url{http://en.wikipedia.org/wiki/Mann-Whitney-Wilcoxon_test}}

We have also tested the hypothesis that for both samples we have the same probability of drawing an author with a larger value of the \textit{collaborative coefficient} \cite[cf.\ also our Table \ref{tab-def}, p.\  \pageref{tab-def}]{Ajiferuke1988collaborative} in the other sample. In both samples we have a similar collaboration behaviour (cf.\ Figure \ref{Fig-coll.coeff}). If we would refuse the null hypothesis we would fail in about one half of possible cases (test probability $p=.516$).
This result ensures that differences between both groups are not due to different typical team sizes. 

All work was done using the free open-source statistics software \textbf{R} (which includes a graphics package).\footnote{~\url{http://www.r-project.org} (\textbf{R}-scripts for indicator calculation and sample data can be obtained from the first author of this paper.)}
\newpage
\subsection{Selection of indicators}

\begin{table}[!b]
\caption{\textbf{List of author indicators}: $a_i$ is the number of authors of paper $i$; $c_i$ is the number of citations of paper $i$; $E(c_i)$ is the expected number of citations of paper $i$ (cf. Appendix \ref{app:exp-cit}); we assume that papers of an author are ordered according to $c_i$ and denote the paper's rank with $r$; the effective rank is defined as $r_\mathrm{eff}(r)= \sum_i^r 1/a_i$.}
\begin{center}
 \begin{tabular}{rl}
 name & definition\\
\hline
\textit{productivity:} &\\
 nr.\ of papers 	& $\sum_i 1=n$\\
 fractional score  	& $\sum_i 1/a_i=f$ \\
\hline
\textit{total influence:}&\\
 nr.\ of citations  	& $\sum_i c_i$\\
norm.\ nr.\ cit. 	& $\sum_i c_i/\mathrm{E}(c_i)$\\
$j$-index		& $\sum_i \sqrt{c_i}$\\
fract.\ citations  	& $\sum_i c_i/a_i$\\
fract.\ norm.\ cit.	& $\sum_i c_i/(\mathrm{E}(c_i)a_i)$\\
\hline
\textit{typical infl.:}&\\
mean cit.\ nr. 	&  $\sum_i c_i/n$\\
mean fract.\ cit.	&  $\sum_i (c_i/a_i)/n$\\
med.\ fract.\ cit.	&  $ \mathrm{median}(c_i/a_i$)\\ 
max. fract.\ cit.	&  $ \max(c_i/a_i)$\\
\hline
\textit{h-type indices:}&\\
Hirsch index  		& $ \max( r | c_r \ge r)$\\
$g$-index		& $ \max( r | \sum_i^r c_i \ge r^2)$\\
\hline
\textit{fract.\ h-type:}&\\
$h_\mathrm{m}$-index  	&  $ \max( r_\mathrm{eff} | c_{r(r_\mathrm{eff}) } \ge r_\mathrm{eff})$\\
$g_\mathrm{f}$-index	&  $ \max( r | \sum_i^{r} c_i/a_i \ge r^2)$\\
$g_\mathrm{m}$-index	&  $ \max( r_\mathrm{eff} | \sum_i^{r(r_\mathrm{eff})} c_i/a_i \ge r_\mathrm{eff}^2)$\\
\hline
\textit{collaboration:}&\\
collab.\ coeff. 		& $1-f/n$\\
\hline
 \end{tabular}
\end{center}
  \label{tab-def}
  \end{table}

The indicators analysed here are listed  together with their mathematical definitions in Table \ref{tab-def}. In Appendix \ref{app:indicators} we discuss the definition 
of each of these indicators. 

We have calculated and tested two simple output indicators and nine indicators of influence.
Beside pure numbers of papers and their citations within the first five publishing years of the authors we use fractionally counted papers and citations as the input for indicators of productivity and of influence. The use of fractional counting in evaluation penalises unjustified assignment of co-authorship to friends.

If we compare papers published in fields with different citation behaviour any citation indicator should be field normalised with expected citation numbers. Here we consider only one field but---as mentioned above---the citation behaviour of astrophysicists has changed dramatically within the last decades. That means,  distributions of unnormalised citation indicators of the two samples overlap  partly due to the changing citation behaviour. 

Another wanted effect of normalising with expected citation numbers is that we account for different citation windows of papers.
Thus, citations to papers published in the beginning of a period obtain a lower weight than those to papers published in the last year. The estimation of expected citation numbers of papers is described in Appendix~\ref{app:exp-cit}.

Another method to deal with varying citation behaviour is to determine each paper's percentile in the citation distribution of a control sample of papers.  \citeN{bornmann_which_2013} compare five approaches to this promising method. Percentile ranking avoids the use of arithmetic means of heavily skewed citation distributions. We minimise the influence of skewness by calculating expected citation numbers by a linear regression over all years considered~(s.~Appendix~\ref{app:exp-cit}).  We have to leave a test of the percentile method with our samples to further work due to a lack of citation data of control samples. 

Recently, several authors tested  a third approach to field normalisation of citation numbers. Here data on the citing side are normalised. \citeN[s.\ also references of this paper]{waltman_systematic_2013} discuss three variants of this method. Also this approach cannot be tested with the data we have at hand.
We could test the simplest variant where each citation of a paper is divided by the number of all references of the citing paper \cite{zhou2011fractional,pepe_measure_2012}. 
\citeN{waltman_systematic_2013} and also \citeN{radicchi_testing_2012}  found that this fractional counting of references does not properly normalise for field and subfield differences.
A further drawback of this variant is that citation numbers are not corrected for the age of the cited paper. We therefore did not test it.

In addition to the eleven indicators of productivity and of influence we calculated the widely used Hirsch or $h$-index~\cite{hirsch2005iqi}, a number combining influence and output performance in an uncontrolled and arbitrary manner, and four variants of it which have been introduced to avoid disadvantages of the Hirsch index.

We did not consider any indicator based on the number of highly cited papers because this contradicts our sampling procedure: we selected \textit{later} stars who have \textit{no} highly cited paper in their first five years of publishing.

\newpage
\section{Results}
Medians of all 16 indicators of both samples are given in Table \ref{tab-medians}. 
In the next to last column of Table~\ref{tab-medians} we list the failure probability $p$ of rejecting the null hypothesis that both samples behave like random samples drawn from the same population.
In the last column we give the rank $R$ according to $p$. For all but the two indicators on least ranks (Hirsch index and \textit{median of fractional citation numbers}) the stars' sample has a higher median than the random sample.  

The boxplots in Appendix \ref{app:boxplots} allow a comparison of indicator distributions for both samples. The figures are ordered according to the ranking $R$. That means that  $p$\,-values increase from the first to the last boxplot. The boxplots have a logarithmic scale because all indicator distributions are highly skewed. All citation indicators have zero values for some uncited authors in the control sample. Therefore we display the logarithm of indicator values~+~1.

The two indicators based on normalised citation numbers are the most useful among the 16 indicators considered (s.~Figure~\ref{Fig-rank1:2}). With respect to \textit{normalised numbers of citations} and to \textit{fractional normalised citations} both samples behave not like random samples from the same population. In both cases, rejecting the null hypothesis has a failure probability below 1\,\%.

\begin{table}[!t]
\caption{Median indicators of samples, test probability $p$, and rank $R$ (according to $p$)}
\begin{center}
 \begin{tabular}{rcccr}
 indicator & stars & random & $p$&$R$\\
\hline
\textit{productivity:} &&&&\\
 nr.\ of papers 	& 8 	& 6 	& .076& 12\\
 fractional score  	& 2.67  & 1.86 	& .095& 13\\
\hline
\textit{total influence:}&&&&\\
 nr.\ of citations  	& 36  	& 22.5 	& .028 & 6\\
norm.\ nr.\ cit. 	& 6.03	& 3.83	& .003 & 1\\
 $j$-index		& 11.86	& 8.76 	& .031 & 9\\
 fract.\ citations  	& 10.00 & 6.57	& .030 & 7\\
 fract.\ norm.\ cit.	& 1.82  & 1.10	& .008 & 2\\
\hline
\textit{typical infl.:}&&&&\\
 mean cit.\ nr. 	&  5.25 & 4.00 	& .117 & 14\\
 mean fract.\ cit.	&  1.23 & 0.99 	& .062 & 11\\
 med.\ fract.\ cit.	&  0.50 & 0.67  & .260 & 16\\
 max. fract.\ cit.	&  4.67 & 3.00 	& .030 & 8\\
\hline
\textit{h-type indices:}&&&&\\
 Hirsch index  		& 3  	& 3 	& .210 & 15\\
 $g$-index		& 5  	& 4	& .037 & 10\\
\hline
\textit{fract.\ h-type:}&&&&\\
 $h_\mathrm{m}$-index  	&  1.32	& 1.00	& .020 & 3\\
$g_\mathrm{f}$-index	&  3 	& 2	& .024 & 4\\
 $g_\mathrm{m}$-index	&  2.38	& 1.68	& .025 & 5\\

\hline
\textit{collaboration:}&&&&\\
collab.\ coeff. 	& .683	& .683	& .516 & 17\\
\hline
 \end{tabular}
\end{center}
  \label{tab-medians}
  \end{table}

The distributions of eight further indicators differ at least on a 5\,\% significance level (s.~Figures \ref{Fig-rank3:4}--\ref{Fig-rank9:10}).   For the remaining six indicators there is no significant difference between distributions of later stars and of authors in the control group~(s.~Figures \ref{Fig-rank11:12}--\ref{Fig-rank15:16}). The Hirsch-index has very similar distributions for both samples~($p=21\,\%$, rank 15, s.\,Figure~\ref{Fig-rank15:16}).
\section{Discussion}

Our results underline the necessity to correct citation indicators for the age of the cited papers and also for varying citation behaviour.\footnote{ It would be interesting---from a theoretical point of view---to determine the influence of each of both corrections separately.} The two indicators of total influence based on citation numbers normalised with expected citation numbers are the only indicators among a total of 16  which show significant differences between later stars and random authors on a 1\,\% level. Thus, normalised citation indicators of total influence can indeed help to predict later successful authors. 
Despite this relative good performance of normalised citation indicators of total influence we cannot recommend to use them as the only basis for an evaluation of young authors in astrophysics and in similar fields of natural sciences. 
Normalisation at the field level cannot correct for a variability in citation numbers between different topics. \citeN{opthof_differences_2011} analysed the citation density in different topics of cardiovascular research papers and concluded that even normalised citation indicators ``should not be used for quality assessment of individual scientists'' (cf.\ his abstract).\footnote{ Topics in physics as in astrophysics also differ substantially in citation density \cite{radicchi_rescaling_2011,pepe_measure_2012}.}
In each case, bibliometrics can only support evaluation and cannot replace individual peer review. 

None of the two output indicators have a significant difference below the 5\,\% level.\footnote{ This is in accordance with the result obtained by \citeN[cf.\,p.\,9]{neufeld_peer_2013} when comparing successful with non-successful applicants of a funding programme for young researchers.} Thus, it is very unlikely to discover a later star in astrophysics by comparing her productivity with the productivity of a random author (Figures~\ref{Fig-rank11:12} and  \ref{Fig-rank13:14}).
The Hirsch index  makes no difference at all~($p = 21\,\%$, Figure~\ref{Fig-rank15:16}). 
This is in agreement with conclusions drawn by 
\citeN{lehmann_measures_2006} and also
by \citeN{kosmulski_calibration_2012} who analysed small samples of mature scientists and found that the number of publications ``is rather useless'' as a tool of assessment and that also the $h$-index is not really helpful.
In contrast to these findings, \citeN{pudovkin_research_2012} found that $h$-index and number of papers are indicators which differ most significantly between group leaders and other scientists at a medical research institution. This  can surely be explained by real output differences of elder and younger researchers but maybe partly also by the assumption that group leaders have more often been  working at the institute over the whole analysed 5-years period than other researchers.

We could have analysed the generalised $h$-index proposed by \citeN{radicchi_universality_2008} who use normalised citation and paper numbers. We did not because $h$ performs much worse than indicators of total influence.

The $g$-index proposed by \citeN{egghe_improvement_2006} to improve the $h$-index  performs indeed better than the original~($p = 3.7\,\%$, Figure~\ref{Fig-rank9:10}). The same holds for the analysed three $h$-type indices which are based on fractional counting. They have been introduced by \citeN{egghe_mathematical_2008} and by  \textcolor{blue}{Schreiber}
 \citeyear{schreiber2008share,schreiber_fractionalized_2009} to account for varying collaboration behaviour. 

There is no significant difference between the two samples when we compare citation  indicators which are designed to reflect the mean influence  of an author's papers. We calculated three of them: 
the arithmetic mean of citation numbers ($p = 11.7\,\%$, Figure \ref{Fig-rank13:14}),  fractionally counted citations per paper ($p = 6.2\,\%$, Figure~\ref{Fig-rank11:12}),
and the median of the fractionally counted citations~($p = 26\,\%$, Figure~\ref{Fig-rank15:16}). 
We wondered whether for a later star a large maximum of (fractional) citations  is more typical than a large value of any measure of central tendency of citation numbers. The answer is yes. The maximum of fractional citations is a better indicator of typical influence ($p = 3\,\%$, Figure~\ref{Fig-rank7:8}). We could have analysed normalised indicators of typical influence, too. We did not because indicators of typical influence do not perform better than those of total influence.


We do not exclude self-citations when calculating citation indicators. There are arguments for their exclusion in evaluative bibliometrics but we assume that it would be difficult for young authors to massively cite their own papers within their first five years of publishing.  

We expect that weighting (fractional) paper numbers with a measure of journal reputation would improve the predictive power of output indicators. We did not test this because the only journal-reputation indicator  available for us was the \textit{journal impact factor} which is not useful here---albeit often used for weighting paper numbers \cite[s.\ also the references of these papers]{Seglen1997impact-factor,lozano_weakening_2012}.

Analysing 85 researchers in oncology \citeN{honekopp2012future} found that ``a linear combination of past productivity and the average paper's citation'' is a better predictor of future publication success than any of the single indicators they had studied. We did not consider combinations of indicators of productivity and of mean influence because the simpler indicators of total influence also reflect productivity---as far as the produced papers have been cited. Neglecting uncited papers is a wanted effect that is also quoted in favour of the $h$-index. 

\citeN{hornbostel_funding_2009} found only small differences in numbers of publications and citations between approved and rejected applicants to a German funding programm for young researchers.
In an earlier study, \citeN{nederhof_peer_1987} compared 19 PhD graduates in physics with best degrees to 119 other graduates with lower grade. They considered the total number of papers before and after graduation and their total and average (short time) impact.  
The 19 best graduates performed significantly better but, interestingly, the impact of their papers declined and reached the level of the control-group papers a few years after graduation. The authors speculate about the reason of this phenomenon and suggest that better students could have been engaged for hot and therefore highly cited research projects. They conclude, that maybe  ``the quality of the research project, and not the quality of the particular graduate is the most important determinant of both productivity and impact figures''~\cite[p.\,348]{nederhof_peer_1987}. This hypothesis could also hold for the young astrophysicists analysed by us. Its confirmation would further diminish the weight of bibliometric indicators in the evaluation of young researchers.   


\section*{Acknowledgements}
We thank Jesper Schneider for helpful discussions of an early draft and Paul Wouters at CWTS in Leiden for providing citation data. The analysis was done for the purposes of the ACUMEN project, financed by the European Commission, cf. \url{http://research-acumen.eu/}.

\newpage

\label{appendix}

\appendix
\section{Appendix}
\subsection{Descriptions of indicators}
\label{app:indicators}
\subsubsection{Productivity indicators}

\paragraph{Number of papers:}
This elementary indicator of productivity belongs to a bygone era when co-authorship was the exception and not the rule. It has the unwanted adverse effects of multiple publishing of the same results and of honorary authorships. 

\paragraph{Fractional score:}
Each paper $i$ is divided into $a_i$ fractions where $a_i$ is the number of authors. These fractions are summed up for the papers  of the evaluated author. We use the simplest variant  where all fractions of a paper are equal: $f = \sum_i 1/a_i$.
This indicator penalises honorary authorships and takes into account that larger teams can be more productive.

\subsubsection{Total influence}
All indicators of total influence tend to increase with the author's number of papers. That means, they are also indicating productivity. 

\paragraph{Number of citations:}
Each citation of a paper indicates that it has influenced the citing author(s). The sum $\sum_i c_i$ of raw numbers $c_i$ of citations of an author's papers is highly field dependent. The paper's number of citations $c_i$ depends on the age of a paper at the time of evaluation. Highly cited papers have surely some quality but less cited ones can also be of high quality.  

\paragraph{Normalised numbers of citations:}
We normalise each paper's number of citations $c_i$ by an expected number of citations $\mathrm{E}(c_i)$ which takes into account the paper's age and the citation behaviour in astrophysics during the first five~(calendar) years in the paper's lifetime (cf.\ Appendix~\ref{app:exp-cit}). After normalising each paper's citation number we sum the ratios of observed and expected citation numbers: $\sum_i c_i/\mathrm{E}(c_i)$.  Some bibliometricians do not calculate the sum of ratios but the ratio of sums  $\sum_i c_i/\sum_i\mathrm{E}(c_i)$~\cite{schubert1986ria}. This procedure is thought to evaluate the whole oeuvre of an author but has been criticised recently for being not ``consistent'' ~\cite{opthof_caveats_2010,waltman_towards_2011}.\footnote{ The $h$-index is also not consistent \cite{marchant_scorebased_2009,waltman2012inconsistency}.} 

\paragraph{The \textit{j}-index:}
The $j$-index is the sum of the square roots of citation numbers of the author's papers $\sum_i \sqrt{c_i}$. It was proposed by \citeN{levene_bibliometric_2012} to downgrade the influence of highly cited papers in the sum of citation numbers.

\paragraph{Fractional citations:}
Analogously to the fractional score described above we distribute citations of each paper equally to its authors: $\sum_i c_i/ a_i.$ 

\paragraph{Fractional normalised citations:}
The normalised numbers of citations can also be  distributed among the authors involved \cite{radicchi_rescaling_2011}:  
$$\sum_{i=1}^n \dfrac{ c_i}{\mathrm{E}(c_i)a_i}.$$

\subsubsection{Typical influence}

\paragraph{Mean citation number:} The arithmetic mean of citations of an author's papers $\sum_i c_i/n$ is the simplest indicator of influence which does not tend to increase with the author's productivity.  

\paragraph{Mean fractional citations:} The arithmetic mean of fractionally counted citations of an author's papers: $\sum_i (c_i/a_i)/n$.  

\paragraph{Median of fractional citations:} The  median of fractionally counted citations of an author's papers $ \mathrm{median}(c_i/a_i)$ is considered because citation distributions are skewed.

\paragraph{Maximum of fractional citations:}
We wondered whether for a later star a large maximum of (fractional) citations $\max(c_i/a_i)$ is more typical than a large value of any measure of central tendency of citation numbers \cite[cf.\ p.\ 375]{lehmann_quantitative_2008}. 

\subsubsection{Indices of \textit{h}-type}
\label{app:h-type}

\paragraph{Hirsch index:} The $h$-index was introduced by~\citeN{hirsch2005iqi} ``to quantify an individual's scientific research output.'' It is defined as the maximum rank $r$ in a rank list of an author's papers according to their citation numbers $c_i$ which is less than or equal to the citation number $c_r$ of the paper with rank $r$: 
$h = \max( r | c_r \ge r).$
The $h$-index has been criticised for its arbitrariness \cite{van_eck_generalizing_2008}. It is arbitrary because in the definition Hirsch ``assumes an equality between incommensurable quantities''~\cite[p.\ 377]{lehmann_quantitative_2008}, namely a rank and a citation number. 
Hirsch himself stated that his index depends on field-specific citation and collaboration behaviour~\cite[p.~16571]{hirsch2005iqi}.  

\paragraph{Egghe's \textit{g}-index:} \citeN{egghe_improvement_2006} criticised the $h$-index for being insensitive to the citation frequency of an author's highly cited papers. His $g$-index can be defined as the maximum rank $r$ which is less than or equal to the mean citation number $(\sum_i^r c_i)/r$ of papers till rank $r$~\cite{schreiber2008influence}. This condition is equivalent to $\sum_i^r c_i \ge r^2$. That means, $g$ can also be defined as
$$ g = \max( r | \sum_{i=1}^r c_i \ge r^2).$$

\subsubsection{Fractional indices of \textit{h}-type}
\label{app:fract-h-type}

\paragraph{Schreiber's $h_\mathrm{m}$-index:}
\label{sec:h_m}
Fractional counting of papers or of citations could be applied to define an $h$-index which takes multi-authorship into account~\cite{egghe_mathematical_2008,schreiber2008share}.
\citeN{schreiber_modification_2008} argued that fractionally counted citations could remove highly cited papers from the $h$-core if they have a lot of authors. This led him to define the $h_\mathrm{m}$-index  
as the maximal effective rank $r_\mathrm{eff}(r)= \sum_i^r 1/a_i$ which is less than or equal to the number of citations $c_r$:	  
$$h_\mathrm{m} = \max( r_\mathrm{eff} | c_{r(r_\mathrm{eff}) } \ge r_\mathrm{eff}).$$

\paragraph{Egghe's $g_\mathrm{f}$-index:}
\label{sec:g_f}
\citeN{egghe_mathematical_2008} proposed to define a fractional $g$-index $g_\mathrm{f}$ as
	  $$g_\mathrm{f} = \max( r | \sum_{i=1}^{r} \frac{c_i}{a_i} \ge r^2).$$
Here the citations are counted fractionally.

\paragraph{Schreiber's $g_\mathrm{m}$-index:}
\label{sec:g_m}
\citeN{schreiber_fractionalized_2009} proposed a fractional $g$-index
$g_\mathrm{m}$ where both, papers and citations, are counted fractionally:	

$$g_\mathrm{m} = \max( r_\mathrm{eff} | \sum_{i=1}^{r(r_\mathrm{eff})} \frac{c_i}{a_i} \ge r_\mathrm{eff}^2).$$

\newpage
\subsection{Expected citation numbers}

\begin{figure}[!b] 
\begin{center} 
\includegraphics[width=2.7in]{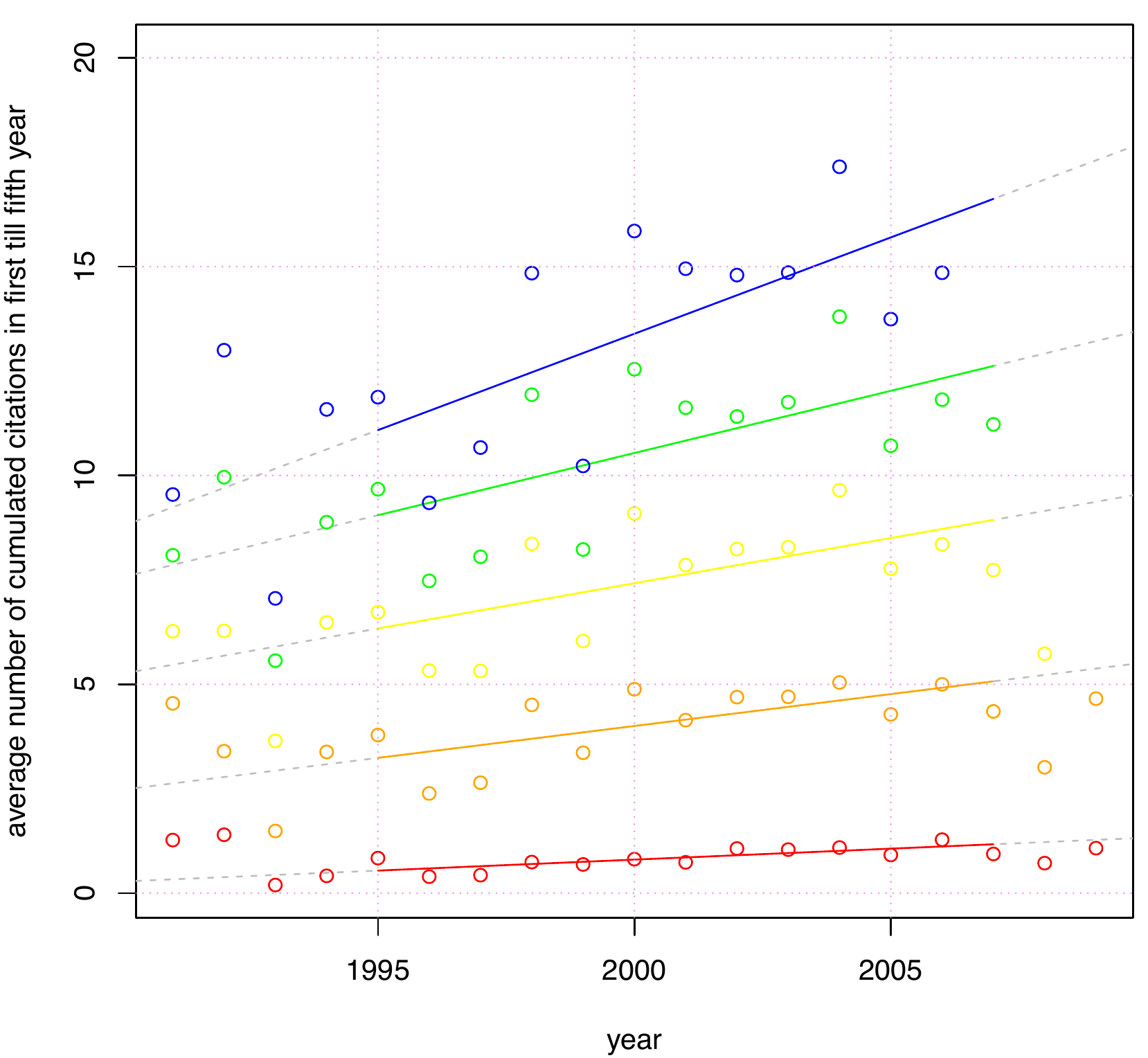} 
\end{center} 
\caption{Linear regressions and averages of citation numbers of papers of random authors in astrophysics after the first (the publication) year (red), the second year (orange), the third year (yellow), the fourth year (green), and the fifth year (blue).} 
\label{Fig-regression-cit} 
\end{figure}
\label{app:exp-cit}
Usually, for field normalisation expected citation numbers of papers are calculated as arithmetic means of citation numbers of all papers (of the same document type) published in all journals of the field in the same year. There are two main technical problems with this method, the rough delineation of fields and the skewness of citation distributions. 

We do not evaluate single authors but only want to show the influence of field normalisation on distributions of citation indicators of authors. Therefore we can use a random sample of papers (for which we have already the citation data) instead of all papers in the field. This sample contains  papers published in the years 1991--2009 by all 331 random authors of our initial control sample. We only consider those 2342 papers with at most 20 authors. Figure \ref{Fig-regression-cit} shows the average cumulated citation numbers in the publication year, one year later, two years later etc. Due to the skewness of citation distributions these arithmetic means fluctuate. Therefore we made a linear regression for each of the five time series of citation numbers of papers (not of the averages) but restricted the analysis to the years 1995--2007 (coloured part of the regression lines) where we have more than 100 papers in each year. The interpolated citation numbers obtained by linear regression are used as expected citation numbers $\mathrm{E}(c_i)$ of papers published in the corresponding years.

From these data we estimate a doubling of citation numbers in astrophysics in the two decades around the millennium.

Calculating expected citation numbers as field averages is problematic because the arithmetic mean is not a good measure for the central tendency of skewed citation distributions. \citeN{lundberg_lifting_2007} therefore proposed to determine expected citation numbers as geometric means of citation numbers of papers in the field. Because papers can have zero citations he adds 1 to be able to calculate the geometric mean. This can be justified by saying that publishing a paper is the first citation of the published results.

\newpage
\subsection{Boxplots of indicators}
\label{app:boxplots}
On this page and the next pages you find boxplots of distributions of all 16 indicators both of the sample of 29 later stars and of the control sample of 74 random young astrophysicists.

\begin{figure}[!b] 
\begin{center} 
\includegraphics[width=3in]{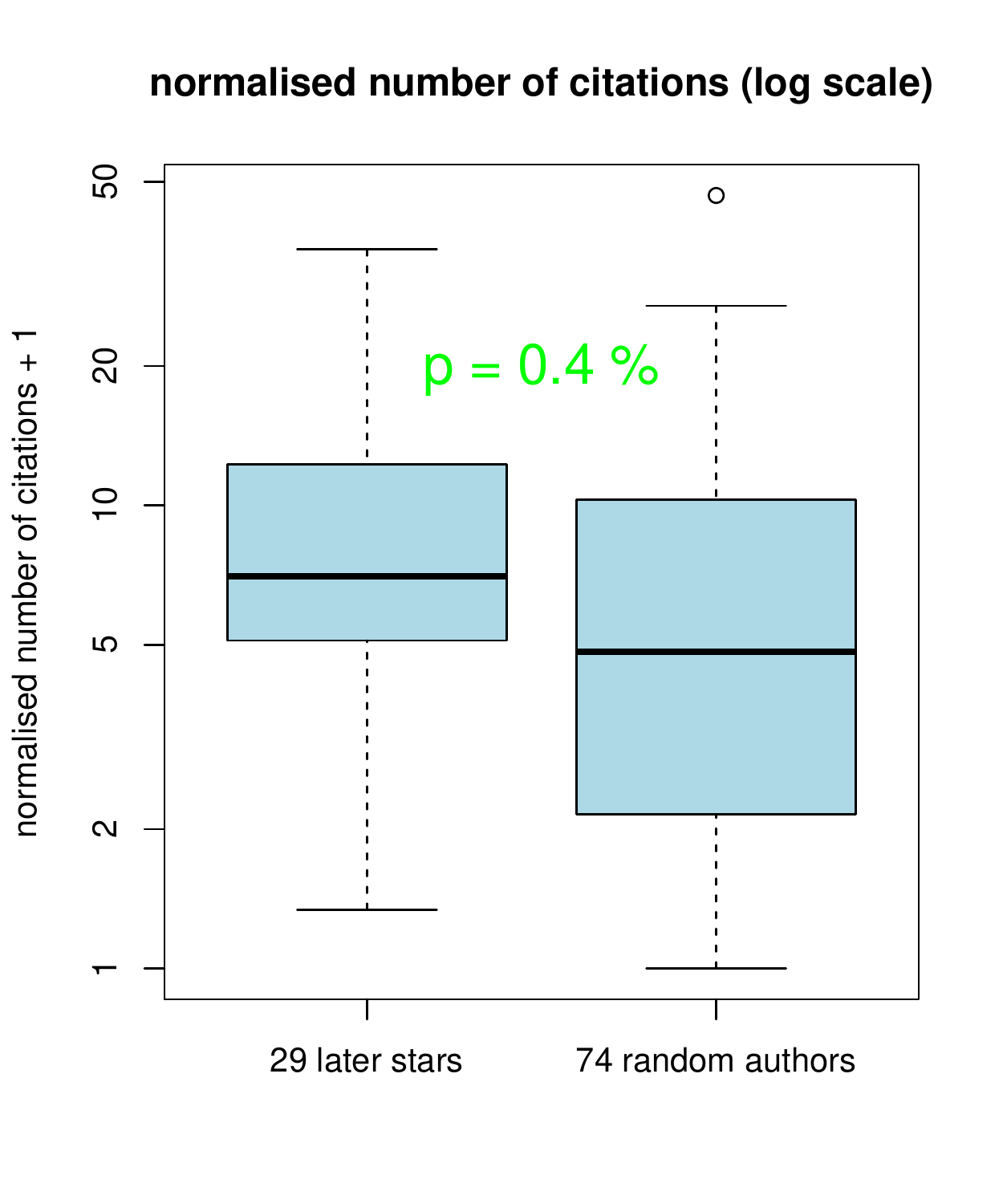} 
\includegraphics[width=3in]{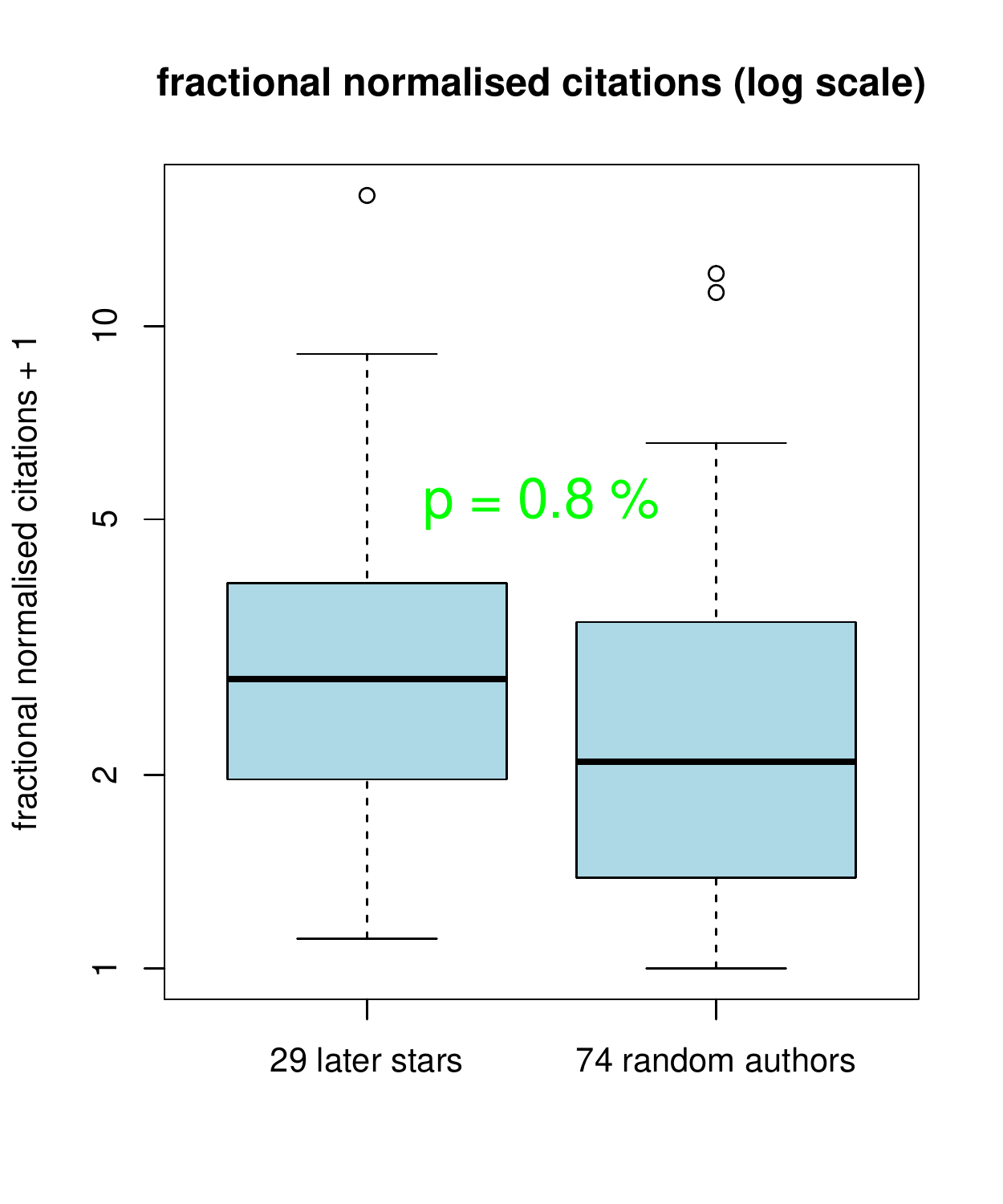} 
\end{center} 
\caption{The two indicators with best $p$-values: $p < 1$\,\% } 
\label{Fig-rank1:2} 
\end{figure}

\begin{figure}[p] 
\begin{center} 
\includegraphics[width=3in]{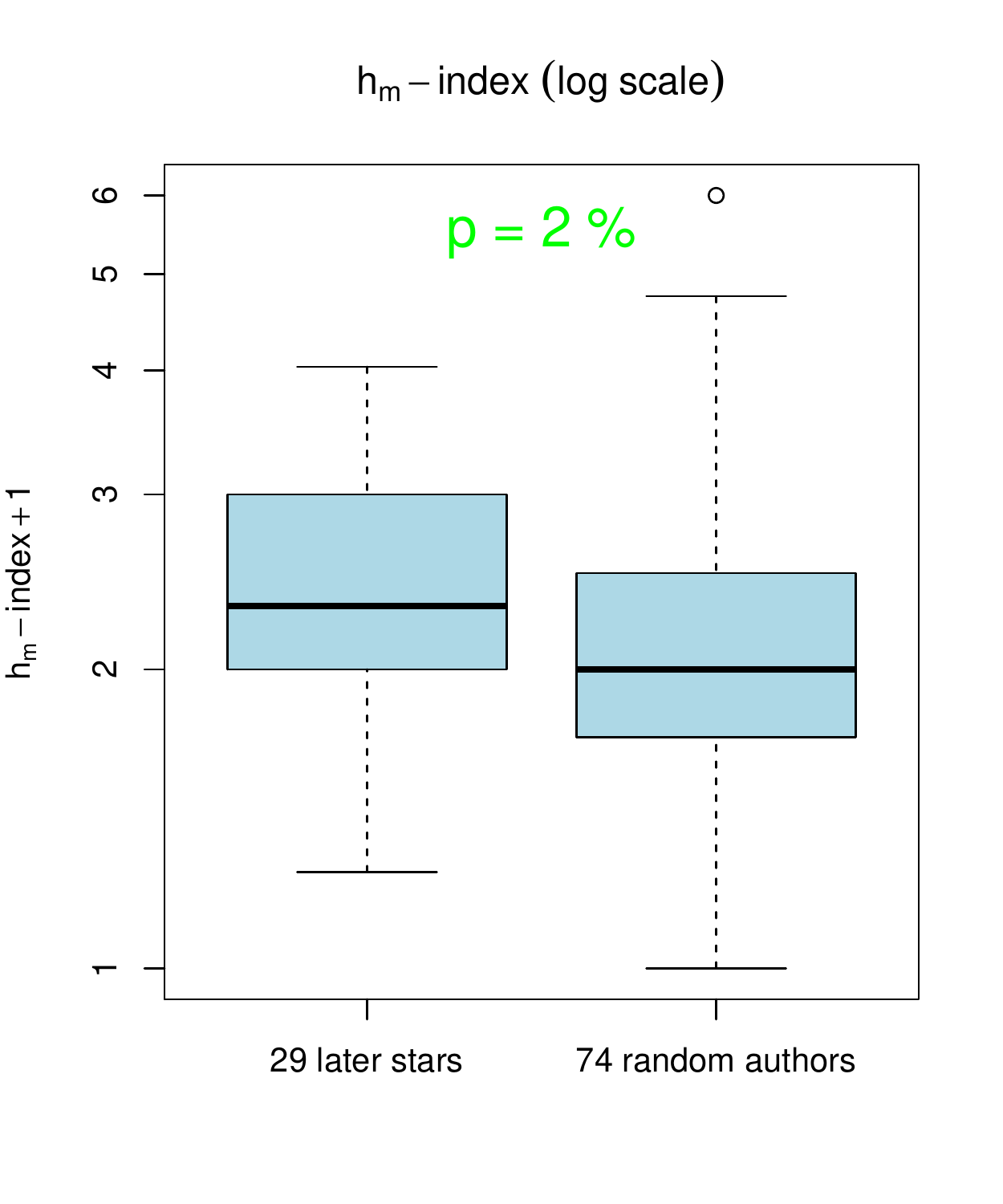} 
\includegraphics[width=3in]{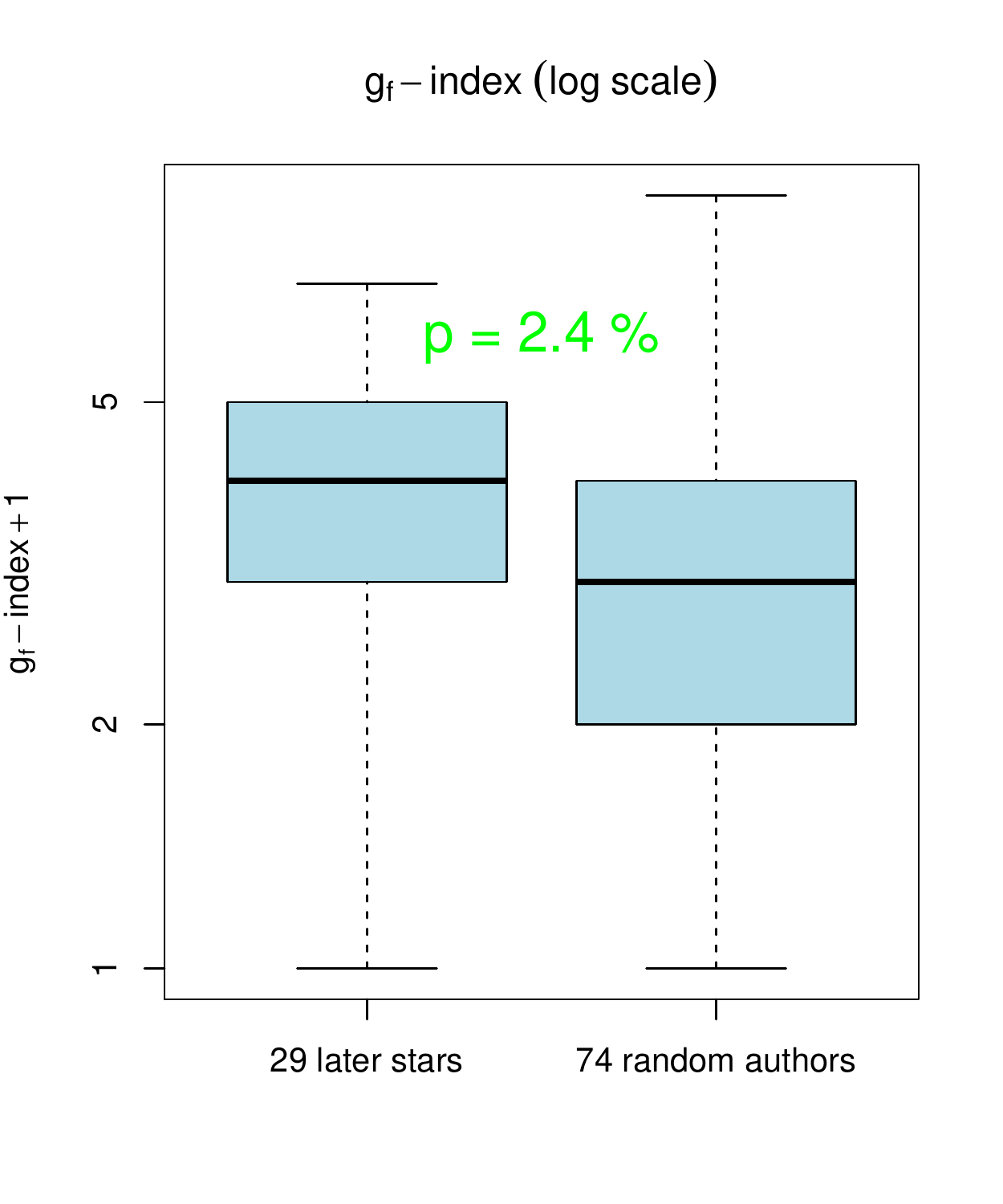} 
\end{center} 
\caption{The indicators on rank 3 and 4 according to $p$-values: $p < 5$\,\%} 
 \label{Fig-rank3:4}
\end{figure}

\begin{figure}[p] 
\begin{center}
\includegraphics[width=3in]{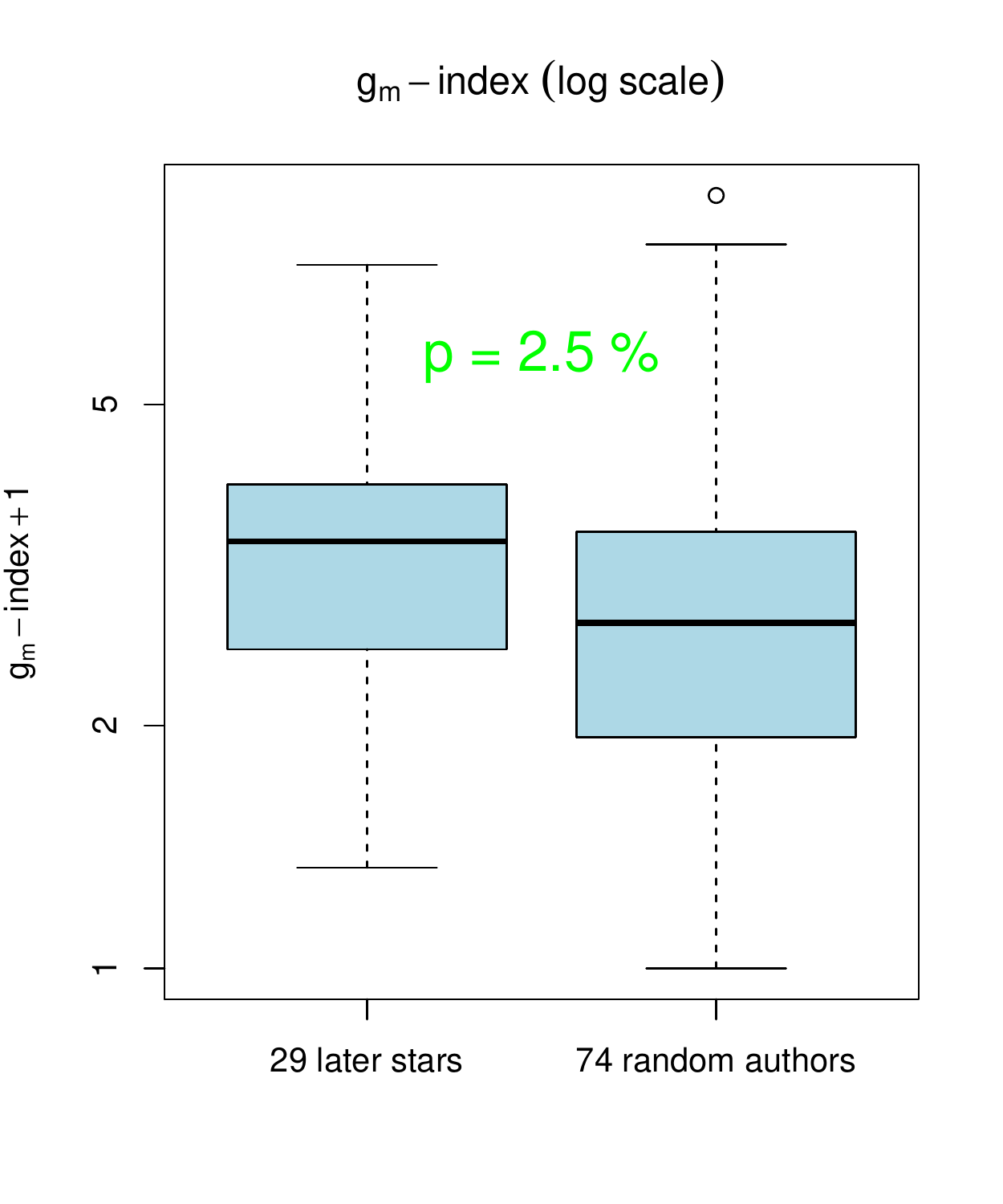} 
\includegraphics[width=3in]{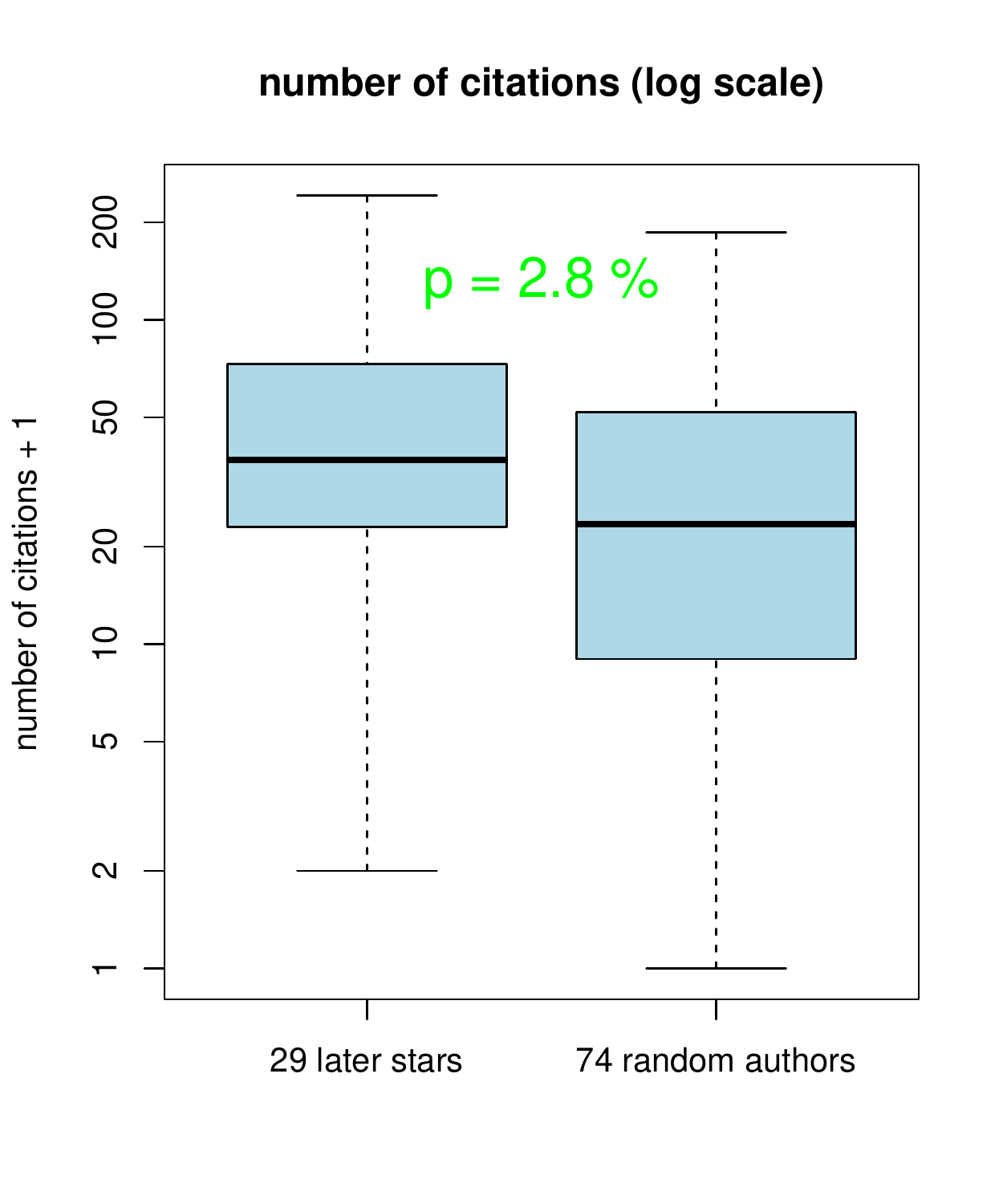} 
\end{center} 
\caption{The indicators on rank 5 and 6 according to $p$-values: $p < 5$\,\%} 
\label{Fig-rank5:6} 
\end{figure}

\begin{figure}[p] 
\begin{center} 
\includegraphics[width=3in]{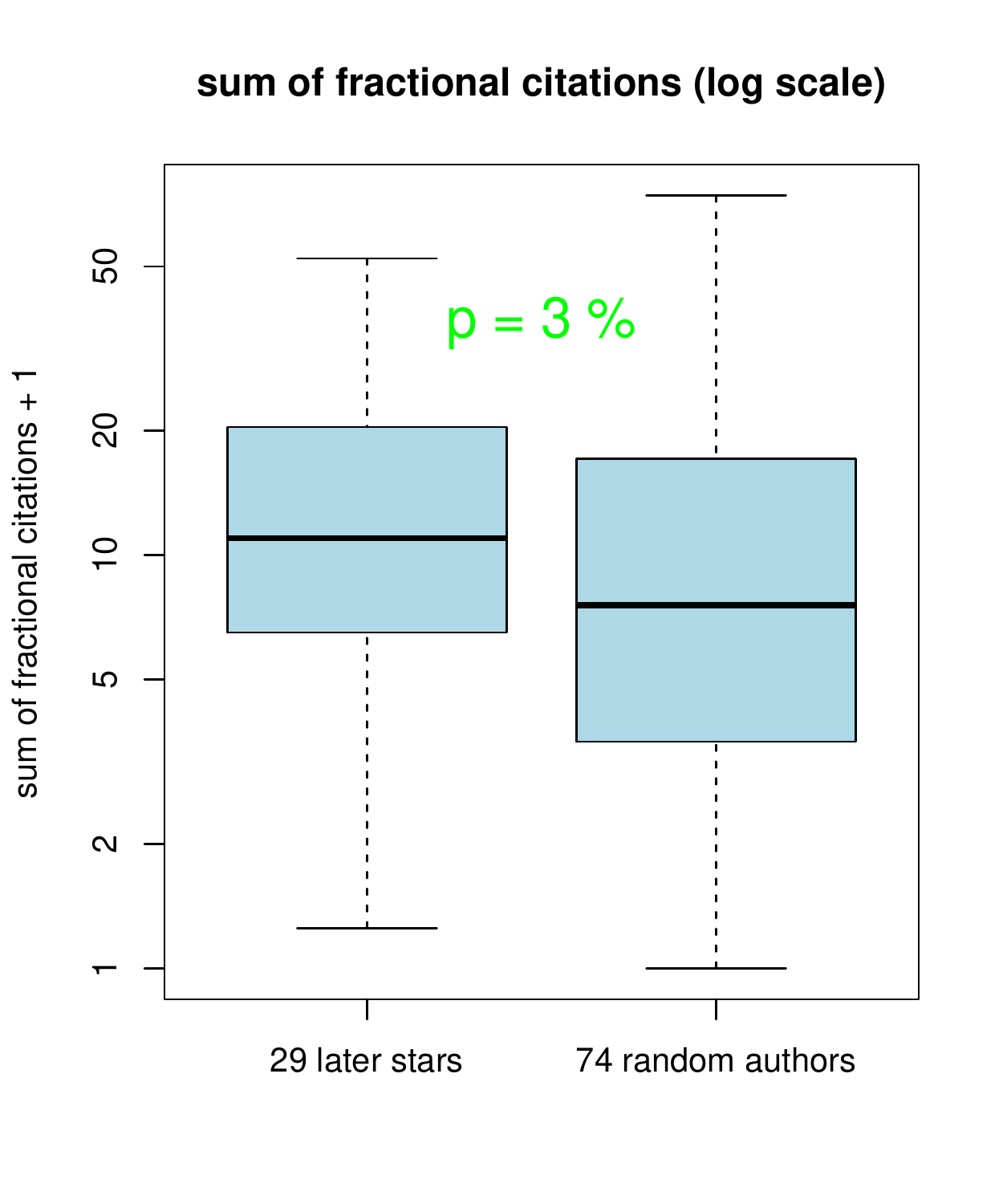}
\includegraphics[width=3in]{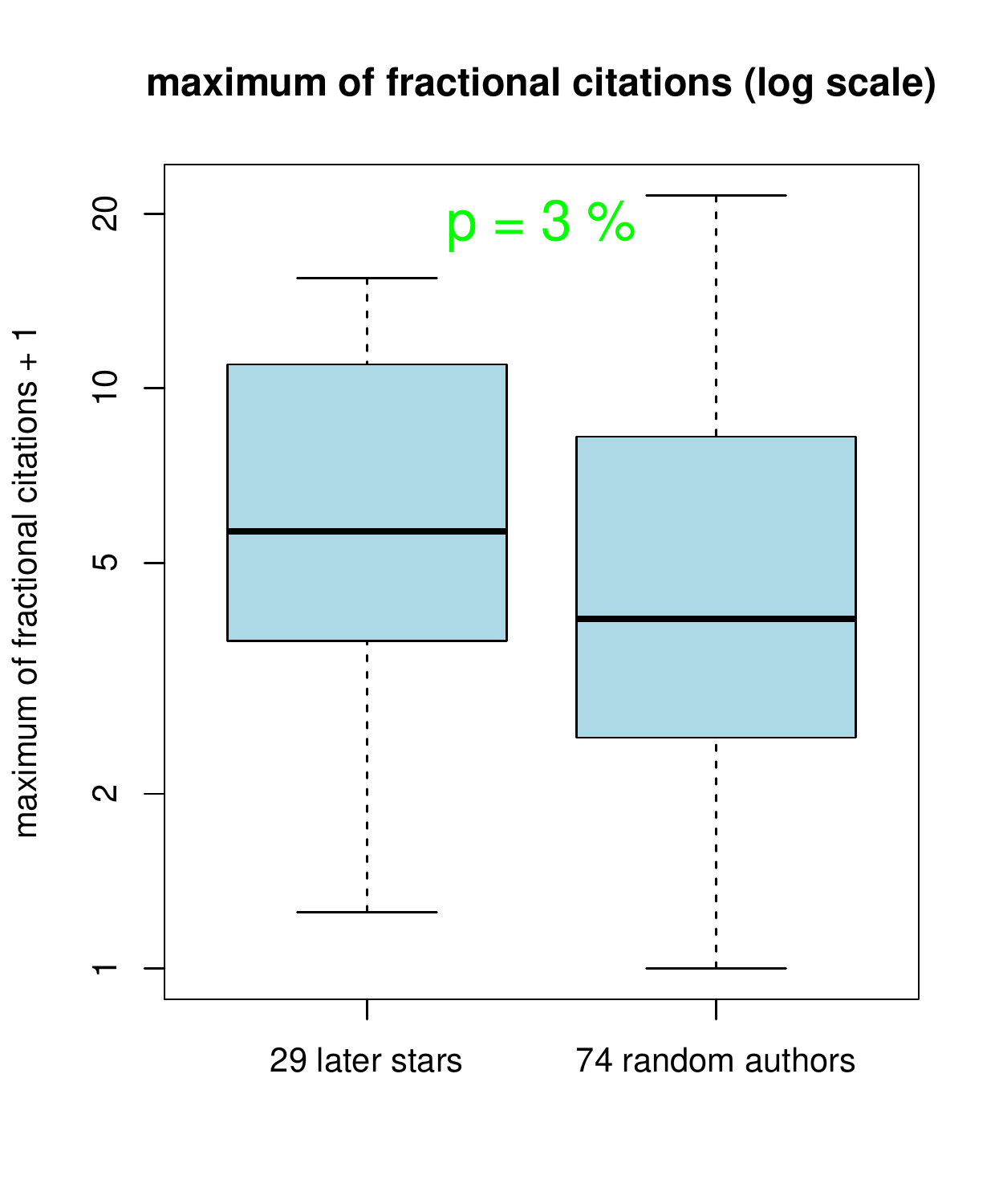}  
\end{center} 
\caption{The indicators on rank 7 and 8 according to $p$-values: $p < 5$\,\%} 
\label{Fig-rank7:8} 
\end{figure}

\begin{figure}[p] 
\begin{center} 
\includegraphics[width=3in]{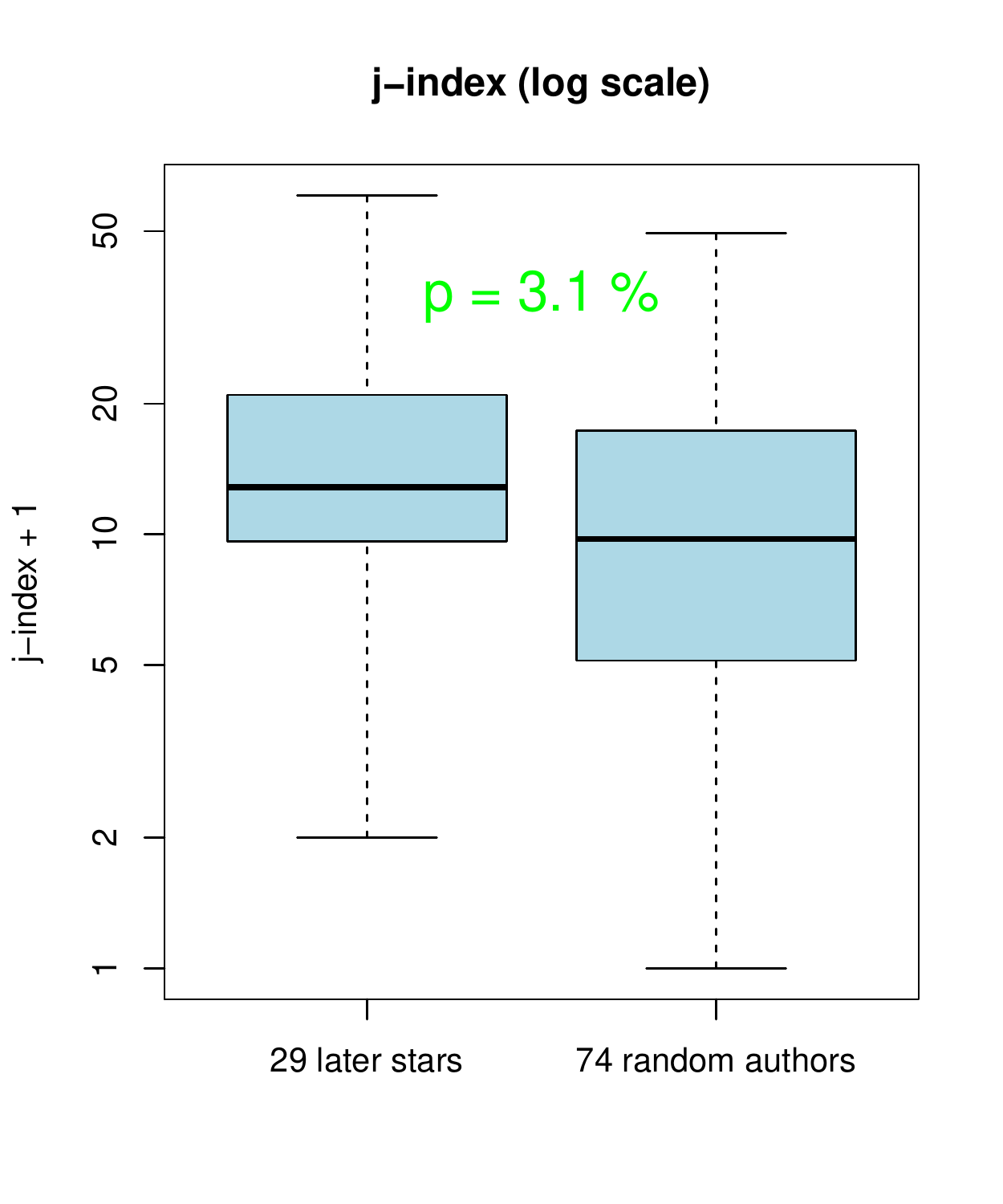}  
\includegraphics[width=3in]{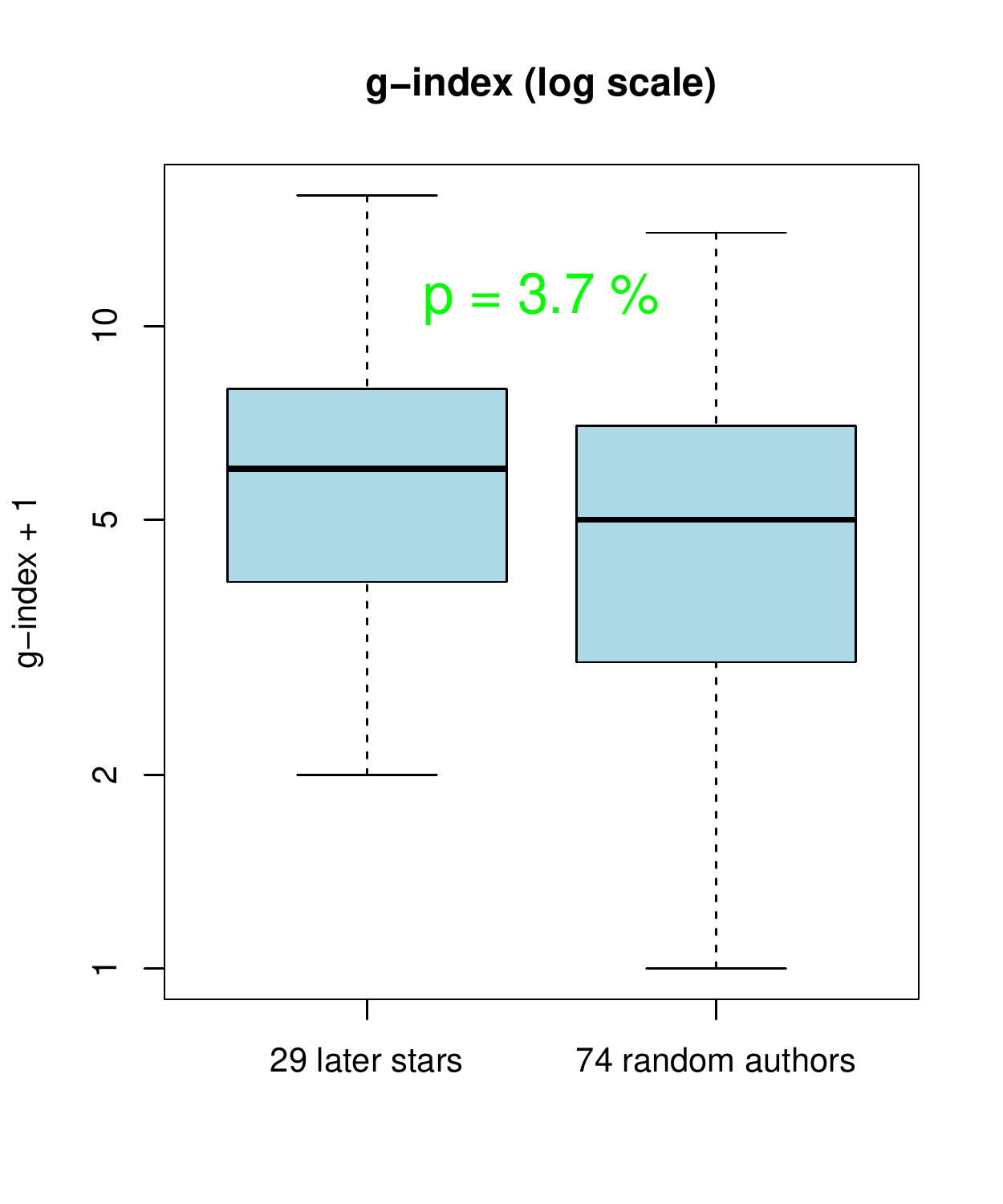}  
\end{center} 
\caption{The indicators on rank 9 and 10 according to $p$-values: $p < 5$\,\%} 
\label{Fig-rank9:10} 
\end{figure}

\begin{figure}[p] 
\begin{center} 
\includegraphics[width=3in]{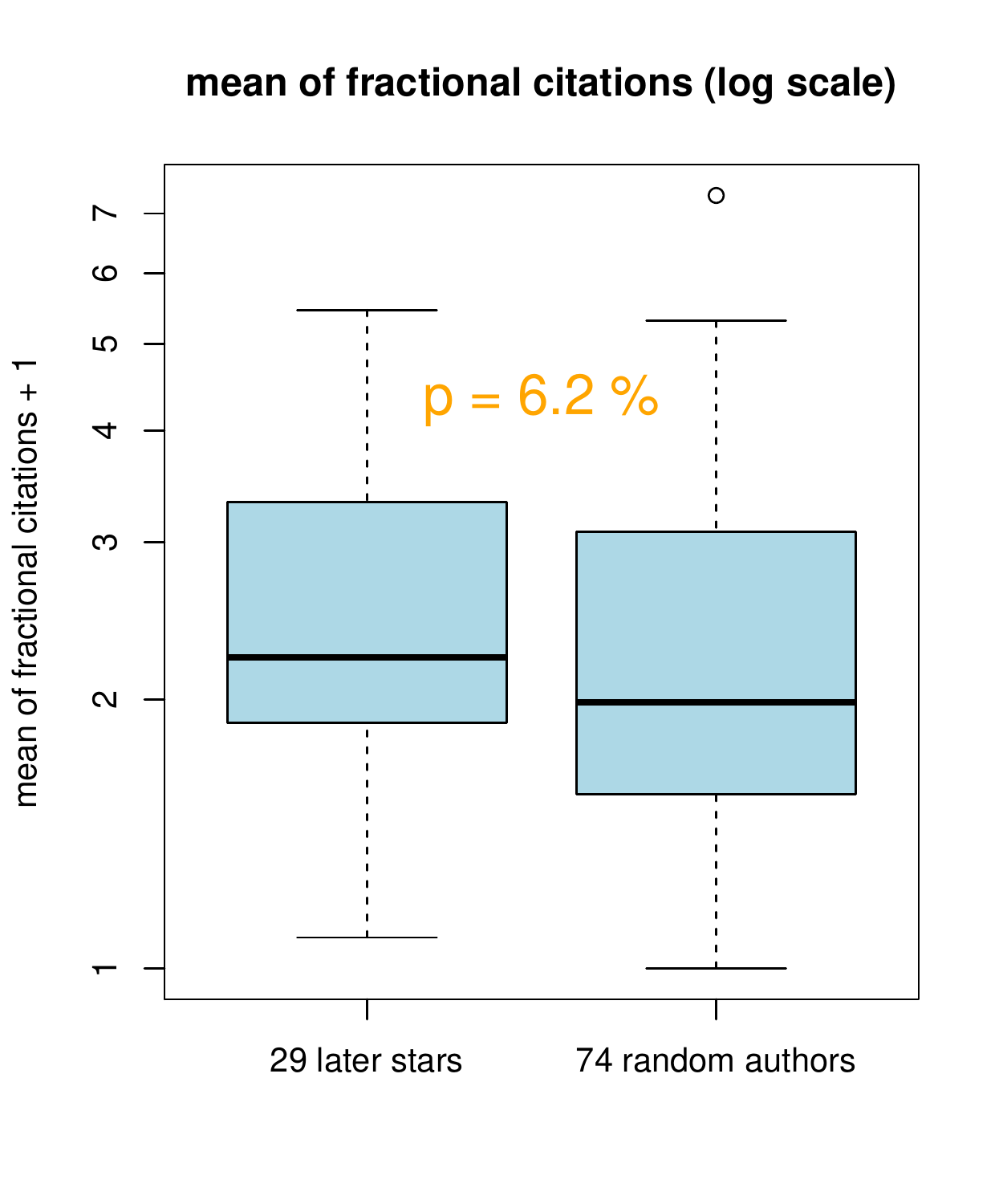} 
\includegraphics[width=3in]{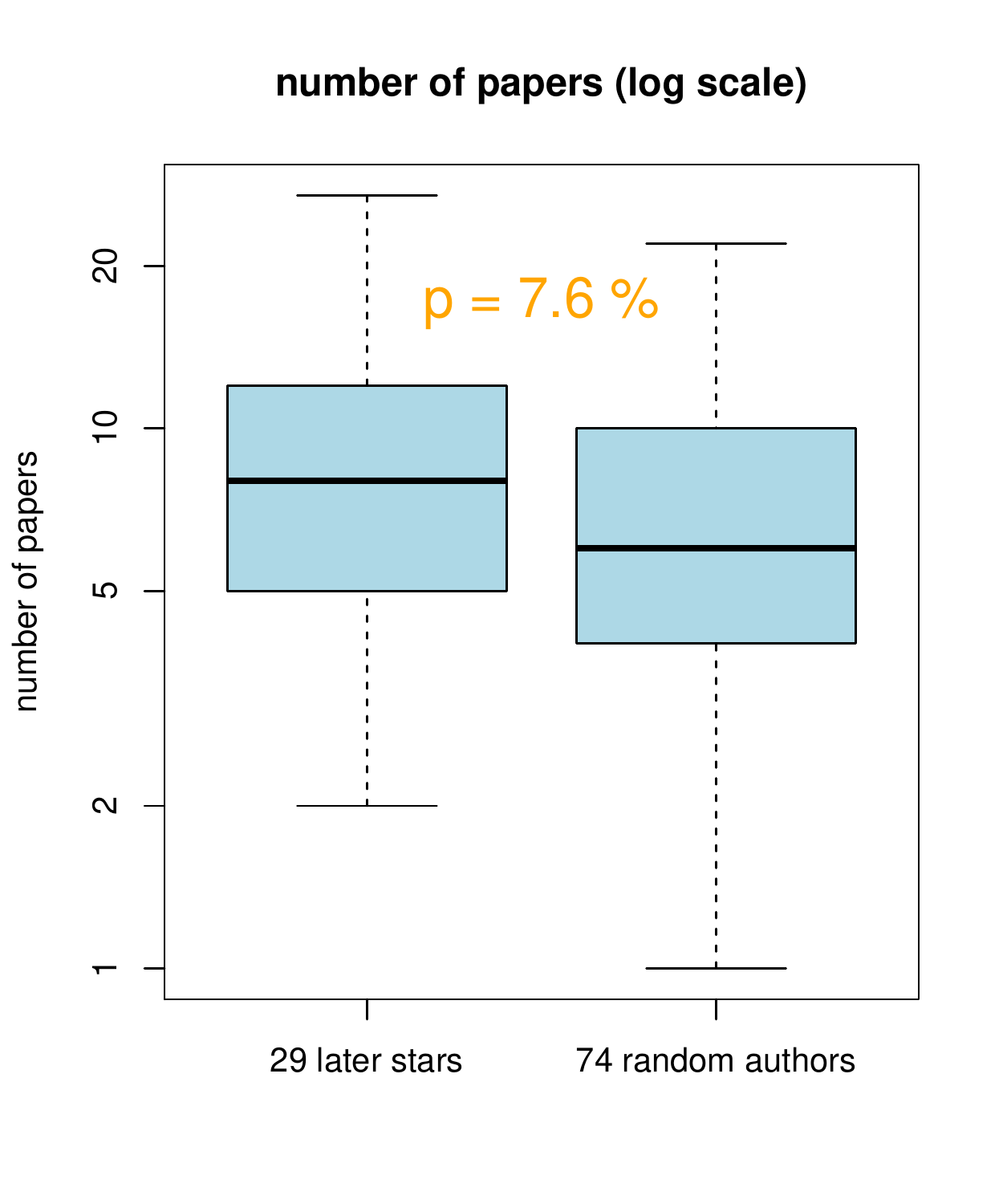} 
\end{center} 
\caption{The indicators on rank 11 and 12 according to $p$-values: $p < 10$\,\%} 
\label{Fig-rank11:12} 
\end{figure}

\begin{figure}[p] 
\begin{center} 
\includegraphics[width=3in]{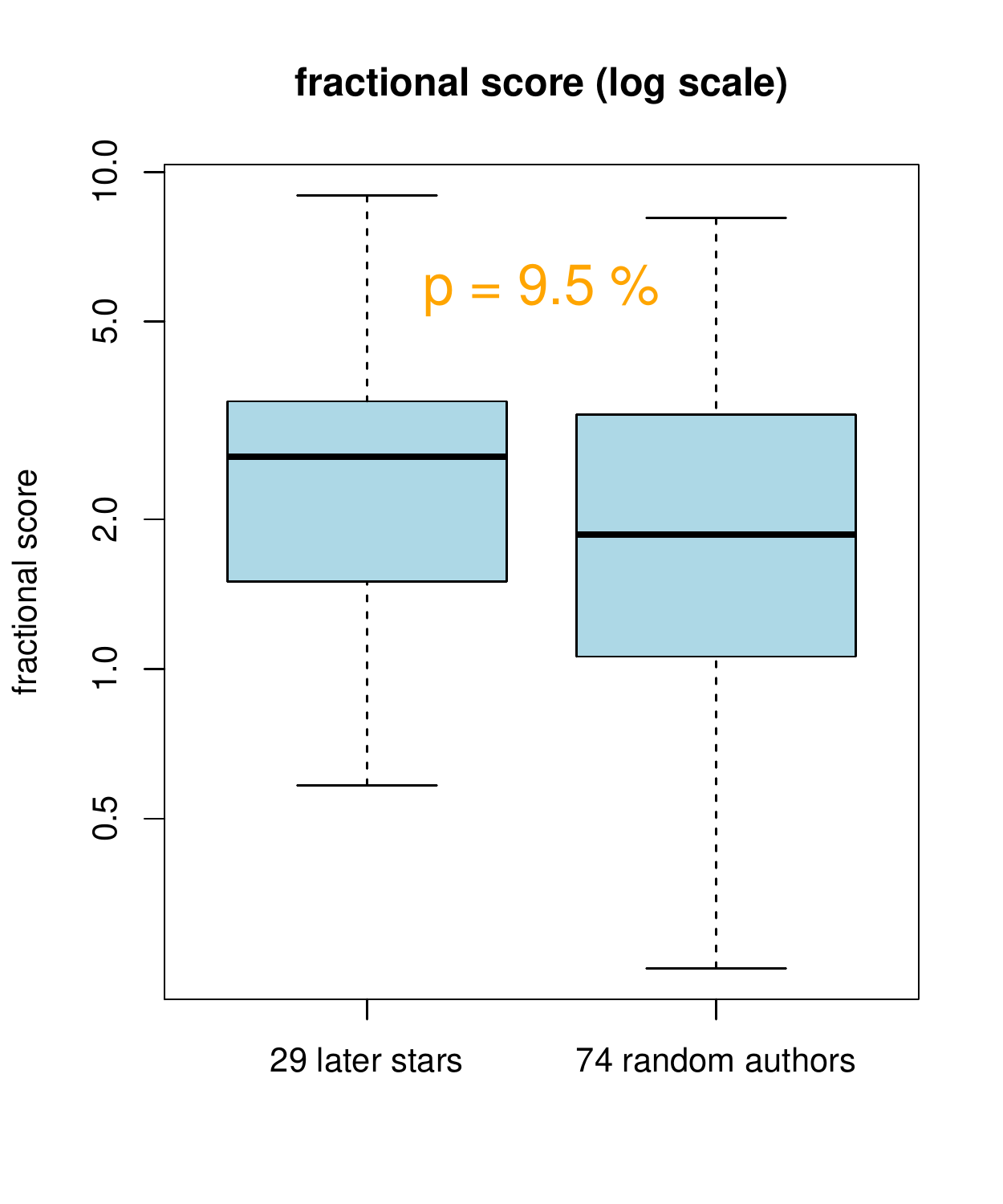} 
\includegraphics[width=3in]{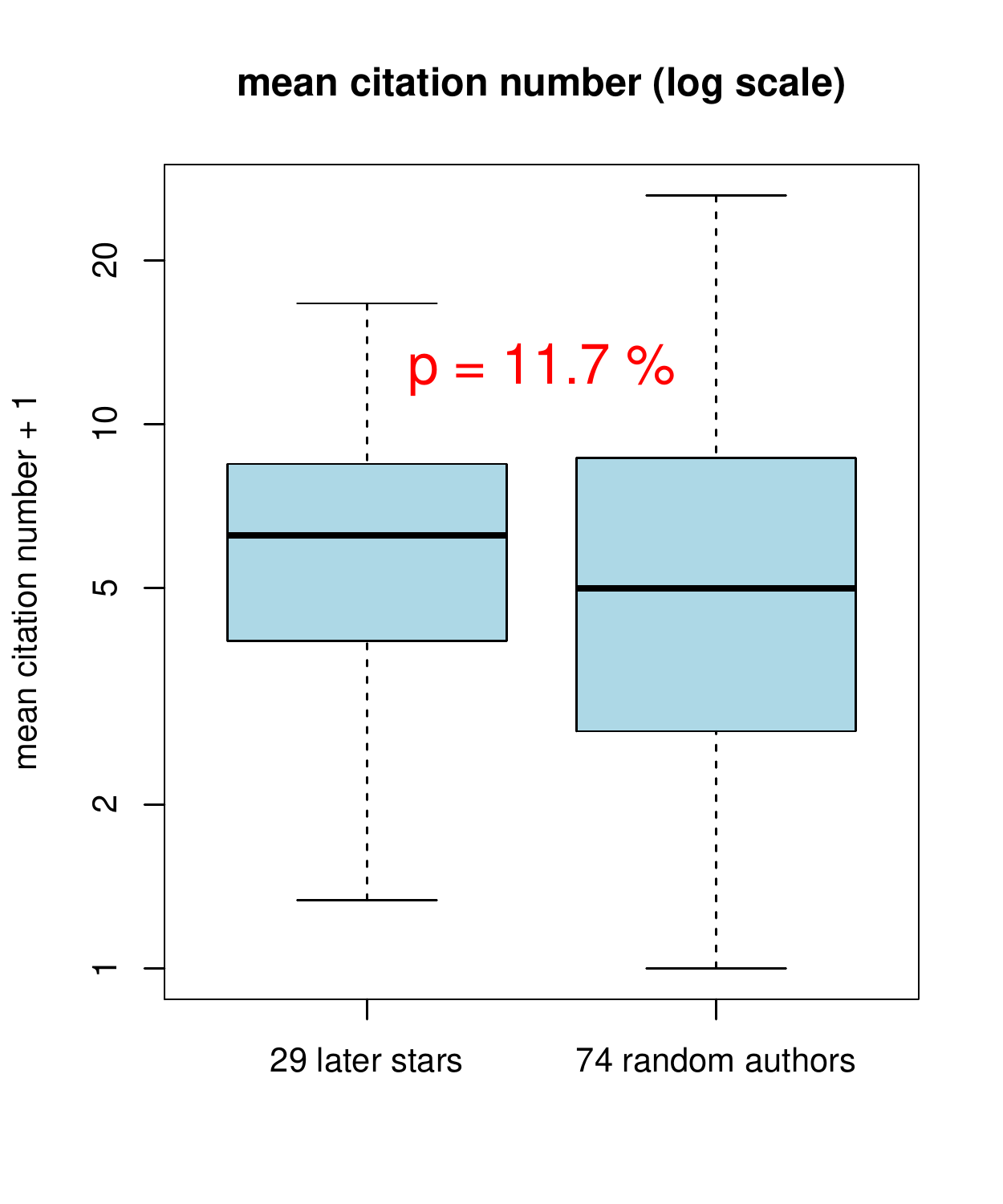} 
\end{center} 
\caption{The indicators on rank 13 and 14 according to $p$-values} 
\label{Fig-rank13:14} 
\end{figure}

\begin{figure}[p] 
\begin{center} 
\includegraphics[width=3in]{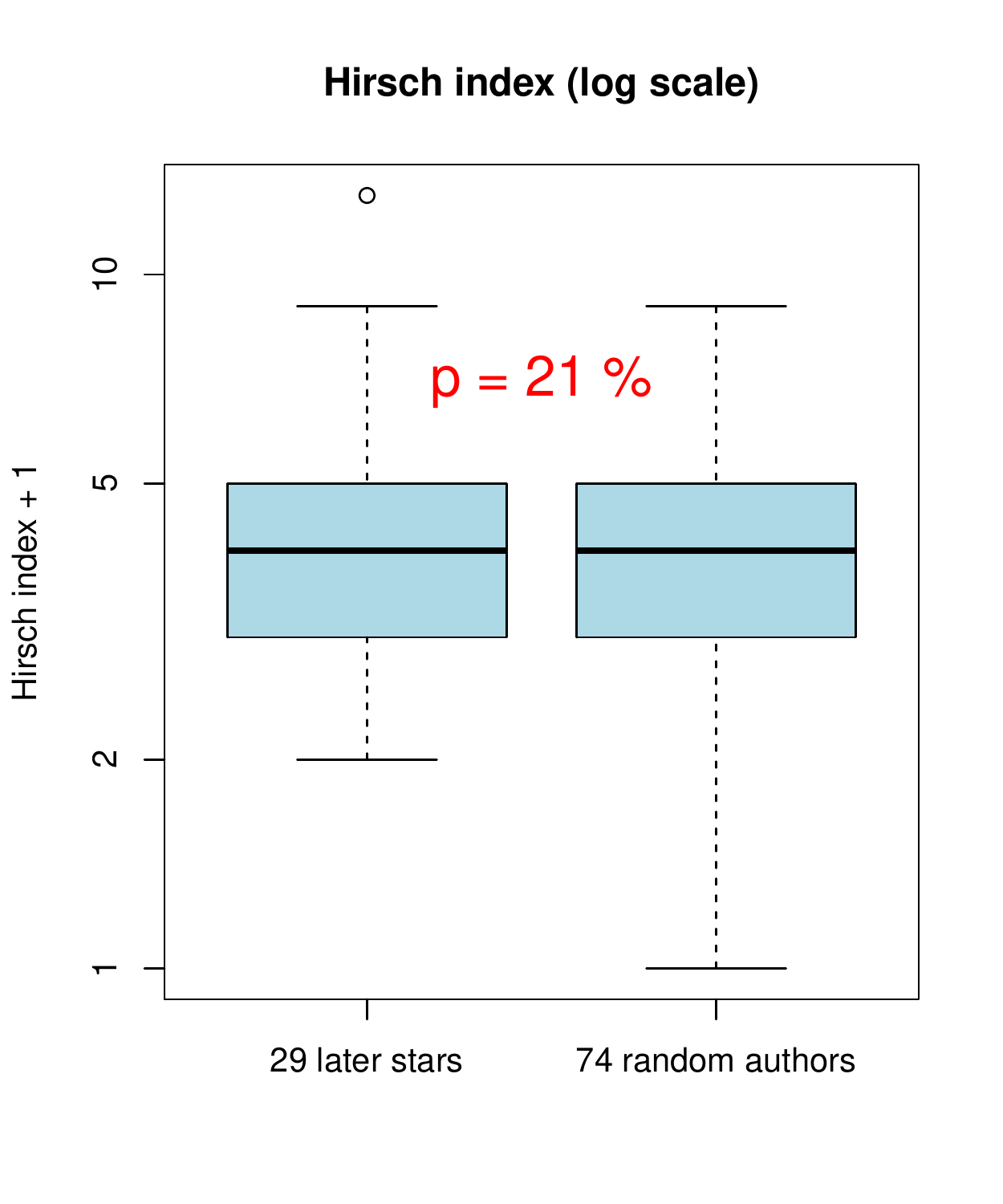} 
\includegraphics[width=3in]{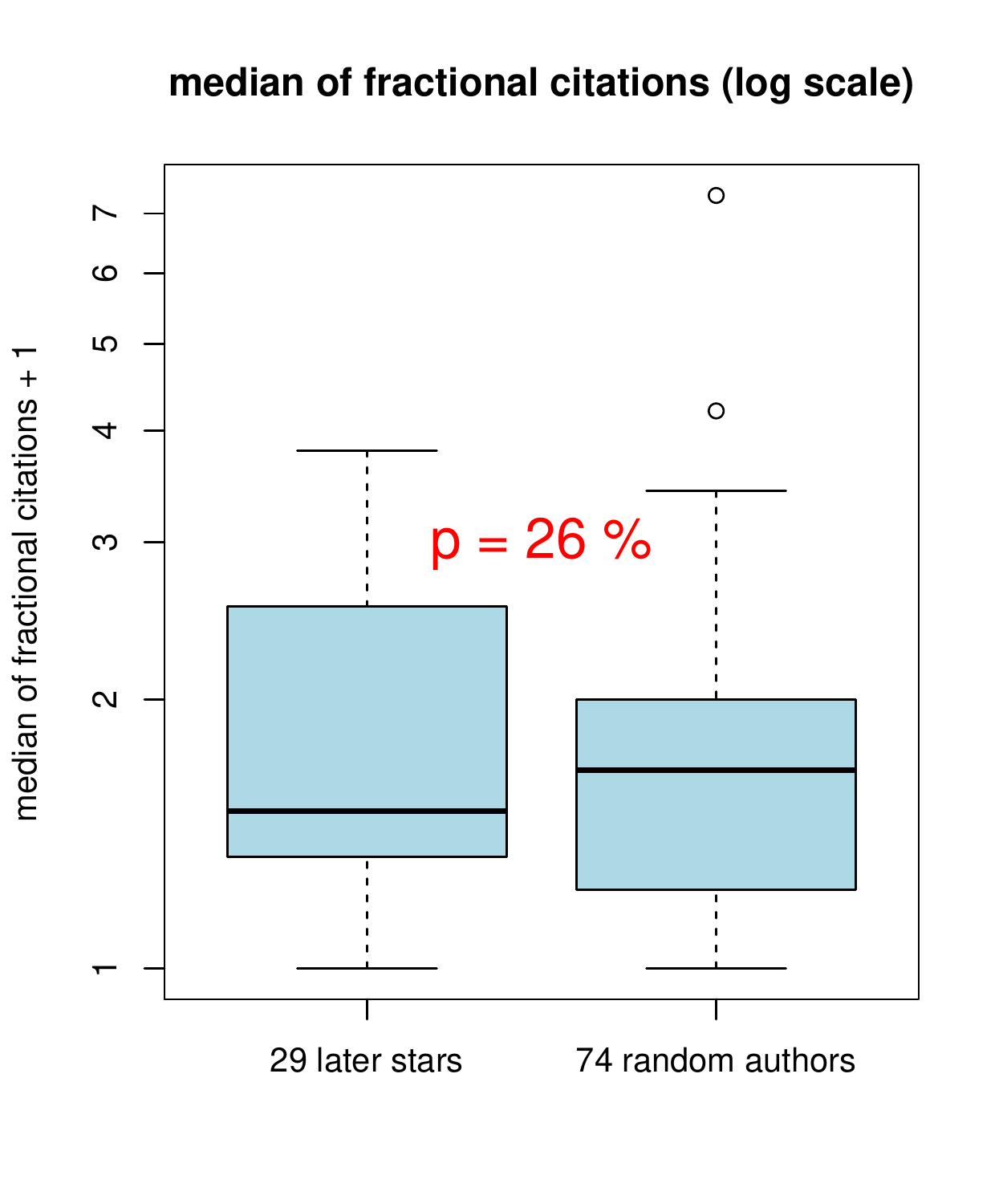} 
\end{center} 
\caption{The indicators on rank 15 and 16 according to $p$-values} 
\label{Fig-rank15:16} 
\end{figure}
\pagebreak
\bibliography{informetrics}

\begin{thebibliography}{}

\bibitem[\protect\citeauthoryear{Ajiferuke, Burrell, and Tague}{Ajiferuke
  et~al.}{1988}]{Ajiferuke1988collaborative}
Ajiferuke, I., Q.~Burrell, and J.~Tague (1988).
\newblock Collaborative coefficient -- a single measure of the degree of
  collaboration in research.
\newblock {\em Scientometrics\/}~{\em 14}, 421--433.

\bibitem[\protect\citeauthoryear{Bornmann, Leydesdorff, and Wang}{Bornmann
  et~al.}{2013}]{bornmann_which_2013}
Bornmann, L., L.~Leydesdorff, and J.~Wang (2013, October).
\newblock {Which percentile-based approach should be preferred for calculating
  normalized citation impact values? An empirical comparison of five approaches
  including a newly developed citation-rank approach (P100)}.
\newblock {\em Journal of Informetrics\/}~{\em 7\/}(4), 933--944.
\newblock s. a. \url{http://arxiv.org/abs/1306.4454}.

\bibitem[\protect\citeauthoryear{Costas, van Leeuwen, and Bordons}{Costas
  et~al.}{2010}]{costas_bibliometric_2010}
Costas, R., T.~N. van Leeuwen, and M.~Bordons (2010, August).
\newblock A bibliometric classificatory approach for the study and assessment
  of research performance at the individual level: The effects of age on
  productivity and impact.
\newblock {\em Journal of the American Society for Information Science and
  Technology\/}~{\em 61\/}(8), 1564--1581.

\bibitem[\protect\citeauthoryear{Egghe}{Egghe}{2006}]{egghe_improvement_2006}
Egghe, L. (2006).
\newblock An improvement of the h-index: the g-index.
\newblock {\em {ISSI} Newsletter\/}~{\em 2\/}(2), 8--9.

\bibitem[\protect\citeauthoryear{Egghe}{Egghe}{2008}]{egghe_mathematical_2008}
Egghe, L. (2008).
\newblock Mathematical theory of the $h$- and $g$-index in case of fractional
  counting of authorship.
\newblock {\em Journal of the American Society for Information Science and
  Technology\/}~{\em 59\/}(10), 1608{\textendash}1616.

\bibitem[\protect\citeauthoryear{Havemann and Larsen}{Havemann and
  Larsen}{2013}]{havemann2013cls}
Havemann, F. and B.~Larsen (2013).
\newblock {Bibliometric Indicators of Young Authors in Astrophysics: Can Later
  Stars be Predicted?}
\newblock In J.~Gorraiz, E.~Schiebel, C.~Gumpenberger, M.~H{\"o}rlesberger, and
  H.~Moed (Eds.), {\em PROCEEDINGS OF ISSI 2013 Vienna}, Volume~2, pp.\
  1881--1883.

\bibitem[\protect\citeauthoryear{Henneken, Kurtz, and Accomazzi}{Henneken
  et~al.}{2011}]{henneken2011ads}
Henneken, E.~A., M.~J. Kurtz, and A.~Accomazzi (2011).
\newblock {The ADS in the Information Age---Impact on Discovery}.
\newblock {\em arXiv preprint arXiv:1106.5644\/}.

\bibitem[\protect\citeauthoryear{Hirsch}{Hirsch}{2005}]{hirsch2005iqi}
Hirsch, J.~E. (2005).
\newblock {An index to quantify an individual's scientific research output}.
\newblock {\em Proceedings of the National Academy of Sciences\/}~{\em
  102\/}(46), 16569--16572.
\newblock \url{http://arxiv. org/abs/physics/0508025}.

\bibitem[\protect\citeauthoryear{H{\"o}ne\-kopp and Khan}{H{\"o}ne\-kopp and
  Khan}{2012}]{honekopp2012future}
H{\"o}ne\-kopp, J. and J.~Khan (2012).
\newblock Future publication success in science is better predicted by
  traditional measures than by the h index.
\newblock {\em Scientometrics\/}~{\em 90\/}(3), 843--853.

\bibitem[\protect\citeauthoryear{Hornbostel, B{\"o}hmer, Klingsporn, Neufeld,
  and von Ins}{Hornbostel et~al.}{2009}]{hornbostel_funding_2009}
Hornbostel, S., S.~B{\"o}hmer, B.~Klingsporn, J.~Neufeld, and M.~von Ins (2009,
  April).
\newblock Funding of young scientist and scientific excellence.
\newblock {\em Scientometrics\/}~{\em 79\/}(1), 171--190.

\bibitem[\protect\citeauthoryear{Kosmulski}{Kosmulski}{2012}]{kosmulski_calibration_2012}
Kosmulski, M. (2012, July).
\newblock Calibration against a reference set: A quantitative approach to
  assessment of the methods of assessment of scientific output.
\newblock {\em Journal of Informetrics\/}~{\em 6\/}(3), 451--456.

\bibitem[\protect\citeauthoryear{Kreiman and Maunsell}{Kreiman and
  Maunsell}{2011}]{kreiman_nine_2011}
Kreiman, G. and J.~H.~R. Maunsell (2011).
\newblock Nine criteria for a measure of scientific output.
\newblock {\em Frontiers in Computational Neuroscience\/}~{\em 5}, article nr.
  48 (6 pages).

\bibitem[\protect\citeauthoryear{Lehmann, Jackson, and Lautrup}{Lehmann
  et~al.}{2006}]{lehmann_measures_2006}
Lehmann, S., A.~D. Jackson, and B.~E. Lautrup (2006, December).
\newblock Measures for measures.
\newblock {\em Nature\/}~{\em 444\/}(7122), 1003--1004.

\bibitem[\protect\citeauthoryear{Lehmann, Jackson, and Lautrup}{Lehmann
  et~al.}{2008}]{lehmann_quantitative_2008}
Lehmann, S., A.~D. Jackson, and B.~E. Lautrup (2008, August).
\newblock A quantitative analysis of indicators of scientific performance.
\newblock {\em Scientometrics\/}~{\em 76\/}(2), 369--390.

\bibitem[\protect\citeauthoryear{Levene, Fenner, and Bar-Ilan}{Levene
  et~al.}{2012}]{levene_bibliometric_2012}
Levene, M., T.~Fenner, and J.~Bar-Ilan (2012).
\newblock A bibliometric index based on the complete list of cited
  publications.
\newblock {\em Cybermetrics: International Journal of Scientometrics,
  Informetrics and Bibliometrics\/}~(16), 1--6.
\newblock s.a.\ arXiv:1304.6945.

\bibitem[\protect\citeauthoryear{Lozano, Larivi{\`e}re, and Gingras}{Lozano
  et~al.}{2012}]{lozano_weakening_2012}
Lozano, G.~A., V.~Larivi{\`e}re, and Y.~Gingras (2012).
\newblock The weakening relationship between the impact factor and papers'
  citations in the digital age.
\newblock {\em Journal of the American Society for Information Science and
  Technology\/}~{\em 63\/}(11), 2140{\textendash}2145.

\bibitem[\protect\citeauthoryear{Lundberg}{Lundberg}{2007}]{lundberg_lifting_2007}
Lundberg, J. (2007, April).
\newblock {Lifting the crown---citation \textit{z}-score}.
\newblock {\em Journal of Informetrics\/}~{\em 1\/}(2), 145--154.

\bibitem[\protect\citeauthoryear{Marchant}{Marchant}{2009}]{marchant_scorebased_2009}
Marchant, T. (2009, June).
\newblock Score-based bibliometric rankings of authors.
\newblock {\em Journal of the American Society for Information Science and
  Technology\/}~{\em 60\/}(6), 1132--1137.

\bibitem[\protect\citeauthoryear{Nederhof and van Raan}{Nederhof and van
  Raan}{1987}]{nederhof_peer_1987}
Nederhof, A.~J. and A.~F.~J. van Raan (1987, May).
\newblock Peer review and bibliometric indicators of scientific performance: A
  comparison of cum laude doctorates with ordinary doctorates in physics.
\newblock {\em Scientometrics\/}~{\em 11\/}(5-6), 333--350.

\bibitem[\protect\citeauthoryear{Neufeld, Huber, and Wegner}{Neufeld
  et~al.}{2013}]{neufeld_peer_2013}
Neufeld, J., N.~Huber, and A.~Wegner (2013, January).
\newblock Peer review-based selection decisions in individual research funding,
  applicants{\textquoteright} publication strategies and performance: The case
  of the {ERC} starting grants.
\newblock {\em Research Evaluation\/}~{\em 22\/}(4), 237--247.

\bibitem[\protect\citeauthoryear{Opthof}{Opthof}{2011}]{opthof_differences_2011}
Opthof, T. (2011, June).
\newblock Differences in citation frequency of clinical and basic science
  papers in cardiovascular research.
\newblock {\em Medical \& Biological Engineering \& Computing\/}~{\em 49\/}(6),
  613--621.

\bibitem[\protect\citeauthoryear{Opthof and Leydesdorff}{Opthof and
  Leydesdorff}{2010}]{opthof_caveats_2010}
Opthof, T. and L.~Leydesdorff (2010, July).
\newblock Caveats for the journal and field normalizations in the {CWTS}
  (``{Leiden}'') evaluations of research performance.
\newblock {\em Journal of Informetrics\/}~{\em 4\/}(3), 423--430.

\bibitem[\protect\citeauthoryear{Pepe and Kurtz}{Pepe and
  Kurtz}{2012}]{pepe_measure_2012}
Pepe, A. and M.~J. Kurtz (2012, November).
\newblock A measure of total research impact independent of time and
  discipline.
\newblock {\em {PLoS} {ONE}\/}~{\em 7\/}(11), e46428.

\bibitem[\protect\citeauthoryear{Pudovkin, Kretschmer, Stegmann, and
  Garfield}{Pudovkin et~al.}{2012}]{pudovkin_research_2012}
Pudovkin, A., H.~Kretschmer, J.~Stegmann, and E.~Garfield (2012).
\newblock {Research evaluation. Part I: productivity and citedness of a German
  medical research institution}.
\newblock {\em Scientometrics\/}~{\em 93\/}(1), 3--16.

\bibitem[\protect\citeauthoryear{Radicchi and Castellano}{Radicchi and
  Castellano}{2011}]{radicchi_rescaling_2011}
Radicchi, F. and C.~Castellano (2011, April).
\newblock Rescaling citations of publications in physics.
\newblock {\em Physical Review E\/}~{\em 83\/}(4), 046116.

\bibitem[\protect\citeauthoryear{Radicchi and Castellano}{Radicchi and
  Castellano}{2012}]{radicchi_testing_2012}
Radicchi, F. and C.~Castellano (2012, January).
\newblock Testing the fairness of citation indicators for comparison across
  scientific domains: The case of fractional citation counts.
\newblock {\em Journal of Informetrics\/}~{\em 6\/}(1), 121--130.

\bibitem[\protect\citeauthoryear{Radicchi, Fortunato, and Castellano}{Radicchi
  et~al.}{2008}]{radicchi_universality_2008}
Radicchi, F., S.~Fortunato, and C.~Castellano (2008, November).
\newblock Universality of citation distributions: Toward an objective measure
  of scientific impact.
\newblock {\em Proceedings of the National Academy of Sciences\/}~{\em
  105\/}(45), 17268--17272.

\bibitem[\protect\citeauthoryear{Schreiber}{Schreiber}{2008a}]{schreiber_modification_2008}
Schreiber, M. (2008a, July).
\newblock {A modification of the $h$-index: The $h_m$-index accounts for
  multi-authored manuscripts}.
\newblock {\em Journal of Informetrics\/}~{\em 2\/}(3), 211--216.

\bibitem[\protect\citeauthoryear{Schreiber}{Schreiber}{2008b}]{schreiber2008influence}
Schreiber, M. (2008b).
\newblock {The influence of self-citation corrections on Egghe's $g$ index}.
\newblock {\em Scientometrics\/}~{\em 76\/}(1), 187--200.
\newblock Cf.\ also arXiv:0707.4577.

\bibitem[\protect\citeauthoryear{Schreiber}{Schreiber}{2008c}]{schreiber2008share}
Schreiber, M. (2008c).
\newblock {To share the fame in a fair way, $h_\mathrm{m}$ modifies $h$ for
  multi-authored manuscripts}.
\newblock {\em New Journal of Physics\/}~{\em 10\/}(4), 040201.

\bibitem[\protect\citeauthoryear{Schreiber}{Schreiber}{2009}]{schreiber_fractionalized_2009}
Schreiber, M. (2009).
\newblock Fractionalized counting of publications for the $g$-index.
\newblock {\em Journal of the American Society for Information Science and
  Technology\/}~{\em 60\/}(10), 2145{\textendash}2150.

\bibitem[\protect\citeauthoryear{Schubert and Braun}{Schubert and
  Braun}{1986}]{schubert1986ria}
Schubert, A. and T.~Braun (1986).
\newblock {Relative indicators and relational charts for comparative assessment
  of publication output and citation impact}.
\newblock {\em Scientometrics\/}~{\em 9\/}(5), 281--291.

\bibitem[\protect\citeauthoryear{Seglen}{Seglen}{1997}]{Seglen1997impact-factor}
Seglen, P.~O. (1997, 2).
\newblock Why the impact factor of journals should not be used for evaluating
  research.
\newblock {\em BMJ: British Medical Journal\/}~{\em 314\/}(7079), 498--513.

\bibitem[\protect\citeauthoryear{van Eck and Waltman}{van Eck and
  Waltman}{2008}]{van_eck_generalizing_2008}
van Eck, N.~J. and L.~Waltman (2008, October).
\newblock Generalizing the $h$- and $g$-indices.
\newblock {\em Journal of Informetrics\/}~{\em 2\/}(4), 263--271.

\bibitem[\protect\citeauthoryear{Waltman and {v}an Eck}{Waltman and {v}an
  Eck}{2012}]{waltman2012inconsistency}
Waltman, L. and N.~J. {v}an Eck (2012).
\newblock {The inconsistency of the \textit{h}-index}.
\newblock {\em Journal of the American Society for Information Science and
  Technology\/}~{\em 63\/}(2), 406--415.

\bibitem[\protect\citeauthoryear{Waltman and van Eck}{Waltman and van
  Eck}{2013}]{waltman_systematic_2013}
Waltman, L. and N.~J. van Eck (2013, October).
\newblock A systematic empirical comparison of different approaches for
  normalizing citation impact indicators.
\newblock {\em Journal of Informetrics\/}~{\em 7\/}(4), 833--849.

\bibitem[\protect\citeauthoryear{Waltman, van Eck, van Leeuwen, Visser, and van
  Raan}{Waltman et~al.}{2011}]{waltman_towards_2011}
Waltman, L., N.~J. van Eck, T.~N. van Leeuwen, M.~S. Visser, and A.~F.~J. van
  Raan (2011, February).
\newblock Towards a new crown indicator: an empirical analysis.
\newblock {\em Scientometrics\/}~{\em 87\/}(3), 467--481.

\bibitem[\protect\citeauthoryear{Zhou and Leydesdorff}{Zhou and
  Leydesdorff}{2011}]{zhou2011fractional}
Zhou, P. and L.~Leydesdorff (2011).
\newblock {Fractional counting of citations in research evaluation: A cross-and
  interdisciplinary assessment of the Tsinghua University in Beijing}.
\newblock {\em Journal of Informetrics\/}~{\em 5\/}(3), 360--368.

\end{thebibliography}
\bibliographystyle{chicago}

\end{document}